\documentclass[ba]{imsart}

\input{preamble}
\newcommand{\nngp}{\textsc{NNGP}}
\newcommand{\qmgp}{\textsc{qmgp}}
\newcommand{\modelname}{GriPS} 
\newcommand{\modelnames}{GriPS}

\pubyear{2025}
\arxiv{2101.03579}
\volume{TBA}
\issue{TBA}
\firstpage{1}
\lastpage{1}

\usepackage{amsthm}
\usepackage{amsmath}
\usepackage{natbib}
\usepackage[colorlinks,citecolor=blue,urlcolor=blue,filecolor=blue,backref=page]{hyperref}
\usepackage{graphicx}
\usepackage{paralist} 

\startlocaldefs
\numberwithin{equation}{section}
\theoremstyle{plain}

\endlocaldefs

\begin{document}

\begin{frontmatter}
\title{Gridding and Parameter Expansion for \\Scalable Latent Gaussian Models of\\ Spatial Multivariate Data}
\runtitle{Gridding and Parameter Expansion}

\begin{aug}
\author{\fnms{Michele} \snm{Peruzzi}\thanksref{addr1}\ead[label=e1]{peruzzi@umich.edu}},
\author{\fnms{Sudipto} \snm{Banerjee}\thanksref{addr2}\ead[label=e2]{sudipto@ucla.edu}},
\author{\fnms{David B.} \snm{Dunson}\thanksref{addr3}\ead[label=e3]{dunson@duke.edu}}
\and
\author{\fnms{Andrew} \snm{Finley}\thanksref{addr4}%
\ead[label=e4]{finleya@msu.edu}}%
\runauthor{M. Peruzzi et al.}

\address[addr1]{Department of Biostatistics, University of Michigan School of Public Health
    \printead{e1} 
}

\address[addr2]{Department of Biostatistics, UCLA Fielding School of Public Health
    \printead{e2}
}

\address[addr3]{Department of Statistical Science, Duke University
    \printead{e3}
}

\address[addr4]{Departments of Forestry and Statistics \& Probability, Michigan State University
    \printead{e4}
}

\end{aug}

\begin{abstract}
Scalable spatial GPs for massive datasets can be built via sparse Directed Acyclic Graphs (DAGs) where a small number of directed edges is sufficient to flexibly characterize spatial dependence. The DAG can be used to devise fast algorithms for posterior sampling of the latent process, but these may exhibit pathological behavior in estimating covariance parameters. 
In this article, we introduce gridding and parameter expansion methods to improve the practical performance of MCMC algorithms in terms of effective sample size per unit time (ESS/s). Gridding is a model-based strategy that reduces the number of expensive operations necessary during MCMC on irregularly spaced data. Parameter expansion reduces dependence in posterior samples in spatial regression for high resolution data. These two strategies lead to computational gains in the big data settings on which we focus. 
We consider popular constructions of univariate spatial processes based on Mat\'ern covariance functions and multivariate coregionalization models for Gaussian outcomes in extensive analyses of synthetic datasets comparing with alternative methods. We demonstrate effectiveness of our proposed methods in a forestry application using remotely sensed data from NASA's Goddard LiDAR, Hyper-Spectral, and Thermal imager (G-LiHT).
\end{abstract}


\end{frontmatter}

\section{Introduction}\label{sec:intro}
Geographic Information Systems (GIS) and related technologies such as remote sensors, satellite imaging and portable devices that collect precise positioning information have led to the amassing of massive amounts of spatial data. This data deluge has motivated substantial developments in modeling and analyzing spatial data using Gaussian Processes (GP), but there remains a clear need for more effective strategies for scaling inference to massive datasets that are now routinely collected. Much of the existing literature on the analysis of massive spatial datasets have devised GPs that scale inference to massive datasets within a Bayesian hierarchical modeling framework \citep[while a comprehensive review is beyond the scope of a single article, see, e.g.,][for in-depth reviews and comparisons of scalable geostatistical methods from different perspectives]{sunligenton, sudipto_ba17, Heaton2019}. These models are widely estimated by simulating the joint posterior distribution of all model parameters and the latent spatial process using customary Markov chain Monte Carlo (MCMC) algorithms. However, insights into efficiently obtaining samples from such high-dimensional posterior distributions in spatial models remain scant. Our current contribution is to achieve computational efficiency in posterior sampling using novel developments in ``gridding'' and ``parameter expansion'' for multivariate latent GP models. 




We specifically focus upon multivariate geostatistics, where we consider measurements on $q$ outcomes with $y_j(\bl)$ denoting the value of the $j$-th outcome, $j=1,\ldots,q$, at a point-referenced spatial location $\bl \in \calD$ within a spatial domain $\calD \subset \Re^{d}$. Each of these measurements is modeled using a spatial regression model,
\begin{equation}
    \label{eq:regression_marginal}
    y_j(\bl) = \bolds{x}_j(\bl)^\top\bbeta_j + w_j(\bl) + \epsilon_j(\bl)\;;\quad j=1,\ldots,q\;,  
\end{equation}
where $\bolds{x}_j(\bl)$ is a $p_j\times 1$ vector of explanatory variables, $w_j(\bl)$ is the latent spatial process associated with outcome $j$ and $\epsilon_j(\bl) \overset{iid}{\sim} N(0,\tau_j^2)$ captures measurement error independently across any set of finite locations. Multivariate spatial analysis builds a $q\times 1$ spatial process $\bw(\bl) = (w_1(\bl), w_2(\bl),\ldots,w_q(\bl))^{\top}$ in a manner that $w_j(\bl)$ captures the spatial association corresponding to outcome $j$, while also capturing associations among the $q$ outcomes. This is achieved by specifying a $q\times q$ matrix-valued \emph{cross-covariance} function $\Cov_{\btheta}(\bl,\bl') = [\text{cov}\{w_i(\bl),w_j(\bl')\}]$ \citep[see, e.g.,][for conditions that ensure valid cross-covariance functions]{genton_ccov}. 
{F}or any positive integer $n$ and any finite collection of points $\calL = \{\bl_1,\bl_2,\ldots,\bl_n\}$%
, %
a $q$-variate multivariate Gaussian process (GP), denoted as $\bw(\bl) \sim GP(\bzero, \Cov_{\btheta}(\cdot,\cdot))$ endows a multivariate Gaussian probability law on any finite collection extracted from $\{\bw(\bl) : \bl \in \calD\}$. Hence, $\bw_{\calL} \sim N(0, \Cov_{\calL})$, where $\bw_{\calL} = (\bw(\bl_1)^{\top},\ldots,\bw(\bl_n)^{\top})^{\top}$ is $nq \times 1$ and $\Cov_{\calL}$ is the $nq \times nq$ block matrix with $q\times q$ cross-covariance $\Cov(\bl_i, \bl_j)$ as its $(i,j)$ block for $i,j=1, \dots, n$. 

We devise methods for efficient Bayesian estimation of multivariate spatial process models such as \eqref{eq:regression_marginal}, where $\bw(\bl)$ scales process-based inference to massive number locations ($n$), but is also endowed with additional structure to achieve dimension reduction when the number of outcomes $q$ is very large. Specifically, we consider cross-covariance structures emanating from the linear model of coregionalization \citep[LMC;][]{wackernagel03, schmidtgelfand}, which produces valid cross-covariance functions for the latent $q$-variate process $\bw(\bl)$. Furthermore, we will focus on spatial factor spatial latent factor models (\citealt{Ren2013, taylor2019spatial, zhangbanerjee20}) that achieve dimension reduction by projecting the $q$-dimensional process into a lower-dimensional subspace spanned by $k < q$ latent GPs. When data size exceeds, e.g., $nq > 10^5$, posterior computations face severe bottlenecks: $\Cov$ is dense and offers no exploitable structure to facilitate computations; the GP produces insurmountable roadblocks in the evaluation of $N(\bw_{\calL}\;; \bzero, \Cov_{\calL})$ as well as in the storage of $\Cov_{\calL}$. 

Our intended contribution here is to devise methods that deliver high efficiency in posterior sampling for (\ref{eq:bayes_hierarchical_lmc}) in terms of the ratio between effective sample size, $T^* < T$, calculated as the size of an i.i.d. sample from the posterior distribution with the same variance as the sample from the Markov chain \citep{liuchen95, robertcasella2004}, and time: $\text{ESS/s} = \frac{T^*}{\text{comp. time}})$. We aim to decrease the denominator and increase the numerator by targeting (i) a reduction in the computational complexity (in terms of floating point operations, or flops, and time) of the algorithm used to generate a sample of size $T$, and (ii) a reduction in posterior dependence of model parameters via reparametrization, leading to an increase in $T^*$. From this perspective, a first step that acts as our baseline for further improvements is to replace the GP with a scalable approximation, 
thereby reducing the time to produce $T$ correlated samples (and thus, $T^*$) from the approximate posterior. In particular, we consider the method of using sparse directed acyclic graphs (DAGs) for introducing large scale spatial conditional independence. Each spatial location (or groups of spatial locations) is a node in the graphical model; other nodes act as parents or children, depending on the direction of the directed edge. If the set of parents is created using neighboring locations, one obtains the popular approximation of \cite{vecchia88}. Subsequent developments \citep[see, e.g.,][]{nngp, guinness_techno, katzfuss_vecchia, meshedgp, spamtrees} extend Vecchia's finite-dimensional approximations to well-defined spatial processes that can be easily incorporated into Bayesian hierarchical models and replace the unrestricted GP, leading to several posterior sampling strategies relying either directly on the sparsity of the graphical model, or on the sparsity of the induced Gaussian precision matrix \citep[see, e.g.,][]{nngp_algos}. 

Although DAG-based methods induce sparsity in 
the precision matrix %
and accrues computational savings in evaluating 
the density of the spatial process, %
such advantages do not necessarily correspond to efficient MCMC algorithms for sampling the posterior distribution (\ref{eq:bayes_hierarchical_lmc}) because high autocorrelations within the Markov chain may require the sampling algorithm be run for very long times. We propose a twofold strategy to improve overall MCMC efficiency. First, we outline gridding as a modeling strategy to reduce the complexity (in terms of flops and time) of each MCMC iteration when using DAG-based GP methods. Second, we consider a form of parameter expansion for covariance parameters with the goal of reducing posterior correlations and improve ESS of the resulting sampler.

The balance of the manuscript proceeds as follows. Section~\ref{sec:multivariate_spatial_regression} provides a brief overview of the class of multivariate spatial models we work with. Gridding and parameter expansion are developed in Sections~\ref{sec:grips:grid}~and~\ref{sec:grips:parametrize}, respectively. Section ~\ref{sec:bayesian_grips_regression} details the resulting regression model. The methods discussed and implemented here are publicly available through the \texttt{meshed} 
software package for the R statistical computing environment, which can be downloaded from the CRAN repository 
at \url{https://cran.r-project.org/package=meshed}.
We illustrate the proposed methodologies in Sections~\ref{sec:applications:simulations:nongridded}~and~\ref{sec:applications:realdata} through simulation experiments and subsequent analysis of next generation Light Detection and Ranging (LiDAR) data from NASA's Goddard LiDAR, Hyper-Spectral, and Thermal imager (G-LiHT) and other remotely sensed variables. 
In the Supplement, we analyze alternative routes such as marginalization of a latent scalable process or the direct approximation of the marginalized model. The Supplement also provides an account of the computational complexity of the proposed methods.

\section{Multivariate spatial regression models: An overview}\label{sec:multivariate_spatial_regression}
We provide some details on the class of multivariate latent Gaussian spatial regression models that form the edifice of our investigations into efficient computation. The model in \eqref{eq:regression_marginal} can be cast into a vector-valued model 
\begin{equation}\label{eq:regression}
\by(\bl) = \bX(\bl)^\top \bbeta + \bw(\bl) + \beps(\bl),
\end{equation}
where $\by(\bl) = (y_1(\bl), y_2(\bl),\ldots,y_q(\bl))^{\top}$ is the $q\times 1$ vector of outcomes at location $\bl$, $\bX(\bl)^\top = \bdiag\{ \bolds{x}_j(\bl)^\top \}_{j=1}^q$ is a $q \times p$, $p=\sum_{j=1}^q p_j$, $\bbeta = (\bbeta_1^{\top},\ldots,\bbeta_{q}^{\top})^{\top}$ is the $p\times 1$ vector of corresponding regression coefficients, $\bw(\bl)$ and $\beps(\bl)$ are $q\times 1$ vectors of spatial random effects and random noise with elements $w_j(\bl)$ and $\epsilon_j(\bl)$, respectively, for $j=1,2,\ldots,q$ such that $\beps(\bl) \iidsim N(0, \bD)$ with a diagonal $\bD = \text{diag}(\tau^2_1, \dots, \tau^2_q)$.

In (\ref{eq:regression}), observations of a continuous multivariate outcome over a finite set of locations are treated as a finite realization of a spatial random field over the domain of interest, plus Gaussian noise. 
We consider multivariate extensions of the Mat\'ern covariance model built using the linear model of coregionalization \citep[LMC;][]{wackernagel03, schmidtgelfand}, which produces valid cross-covariance functions for the latent $q$-variate process $\bw(\cdot)$ in (\ref{eq:regression}). The LMC-Mat\'ern is particularly convenient as it generalizes the ubiquitous univariate Mat\'ern model to multivariate settings. In the LMC-Mat\'ern model, $\bw(\bl)$ can be expressed as
\begin{equation}\label{eq:lmc}
\begin{aligned}
    \bw(\bl) &= \sum_{j=1}^k \blambda_j \omega_j(\bl) = \begin{bmatrix} 
\lambda_{11} \omega_1(\bl) + \dots + \lambda_{1k} \omega_k(\bl) \\
\vdots \\
\lambda_{q1} \omega_1(\bl) + \dots + \lambda_{qk} \omega_k(\bl)
\end{bmatrix} = \bLambda \bomega(\bl)
\end{aligned}
\end{equation}
where $\bLambda = [\blambda_1,\ldots,\blambda_k]$ is $q \times k$ with rank $k$ and $(i,j)$th entry $\lambda_{ij}$. We specify (\ref{eq:regression}) as
\begin{equation}\label{eq:lmc_regression}
    \by(\bl) = \bX(\bl)^\top \bbeta + \bLambda \bomega(\bl) + \beps(\bl)\;,\\
\end{equation}
where we have made the LMC model explicit. Each $\omega_j(\bl)$ is an independent univariate spatial process with Mat\'ern correlation $\rho_j(\bl, \bl')$, where
\begin{align}\label{eq:matern}
   \rho_j(\bl, \bl') = \rho(\bl, \bl'; \{\phi_j, \nu_j\}) &= \frac{2^{1-\nu_j}}{\Gamma(\nu_j)} \phi_j^{\nu_j} \| \bl - \bl' \|^{\nu_j} K_{\nu_j}\left( \phi_j \| \bl - \bl' \|\right),
\end{align}
and $K_{\nu_j}$ is the modified Bessel function of the second kind of order $\nu_j$. In the following, we will take $\nu_j = \nu$ for all $j$ and consider it known; therefore, $\btheta = (\bLambda, \bphi)$. The $k\leq q$ components of $\bomega(\bl)$ are independent of each other (hence $\text{cov}\{\omega_j(\bl), \omega_h(\bl') \} = 0$ whenever $h\neq j$). Hence, $\bomega(\bl)$ is a multivariate spatial process with cross-correlation $\brho(\bl, \bl'; \bvarphi)$ indexed by $(\bvarphi, \nu) = \{\phi_1, \dots, \phi_k, \nu\}$. The cross-covariance function induced by the LMC-Mat\'ern model is thus $\Cov_{\btheta}(\bl, \bl') = \bLambda \brho(\bl, \bl'; \bphi) \bLambda^\top$ and satisfies (i) $\Cov_{\btheta}(\bl,\bl') = \Cov_{\btheta}(\bl',\bl)^{\top}$; and (ii) $\sum_{i=1}^n\sum_{j=1}^n \bz_i^{\top}\Cov_{\btheta}(\bl_i,\bl_j)\bz_j > 0$ for any integer $n$ and any finite collection of points $\{\bl_1,\bl_2,\ldots,\bl_n\}$ and for all $\bz_i \in \Re^{q}\setminus \{\bolds{0}\}$. 
If $q > k$, one obtains a spatial latent factor model. We will henceforth omit $\btheta$ from notation for simplicity.

Letting $\calT = \{ \bl_1, \dots, \bl_n \} \subset \calD$ be the set of locations where at least one of the $q$ spatial outcomes has been measured, we construct $\by = [\by(\bl_1)^\top, \dots, \by(\bl_n)^\top]^\top$ as the $nq\times 1$ vector of $\by(\bl_i)$'s over the $n$ locations, analogously define $\bw$ and $\beps$, and let $\Cov=\Cov_{\calT}$ and $\bX = [\bX(\bl_1) : \ldots : \bX(\bl_n) ]^\top$. We also denote $\btau = \{\tau_1^2, \dots, \tau_q^2\}$, and introduce $\bolds{o}_\calT$ as the binary vector whose zero entries correspond to unobserved elements of $\by$, $\bD_n = \bdiag(\{ \bD \}_{i=1}^n)$, and let $\odot$ denote the Hadamard element-by-element product. After endowing all unknown parameters in (\ref{eq:regression}) with prior distributions, the resulting Bayesian hierarchical model leads to the posterior distribution 
\begin{equation}\label{eq:bayes_hierarchical_lmc}
\begin{aligned}
    &p(\bbeta,\btau,\bomega,\bLambda,\bvarphi\given \by) = p(\bbeta,\btau,\bw,\btheta \given \by)\\
    &\qquad \propto  p(\bbeta, \btau, \btheta) \times N(\bw ; \bzero, \Cov) \times N(\by \odot \bolds{o}_\calT \;; (\bX\bbeta + (\bI_n \otimes \bLambda) \bomega) \odot \bolds{o}_\calT, \bD_n)\;.
\end{aligned}
\end{equation}
This fully specifies a Bayesian hierarchical model with a latent spatial process embedded within. Such models render greater flexibility in many realistic scenarios. For example, they easily accommodate analysis of non-Gaussian data. In multivariate settings, latent models straightforwardly enable Bayesian analyses of multi-type or misaligned outcomes. 

In order to scale inference to massive datasets, we replace $\Cov$ with a covariance matrix $\tilde{\Cov}$ such that $N(\bw; \bzero, \tilde{\Cov}) \approx N(\bw; \bzero, \Cov)$ is computationally efficient. One approach is to use a DAG to construct the approximate spatial covariance matrix $\tilde{\Cov}$ such that the precision, $\tilde{\Cov}^{-1}$ is sparse. While a substantial literature on DAG-based methods have illustrated their effectiveness in scaling inference to massive datasets by exploiting cheap storage and computation of $\tilde{\Cov}^{-1}$,  such advantages do not necessarily correspond to efficient MCMC algorithms for sampling the posterior distribution (\ref{eq:bayes_hierarchical_lmc}) because high autocorrelations within the Markov chain may require the sampling algorithm be run for very long times. Our twofold strategy to improve efficiency of posterior sampling consists of two ideas. First, we introduce gridding (Section~\ref{sec:grips:grid}) as a modeling strategy to reduce the complexity of each MCMC iteration. Second, we consider a form of parameter expansion (Section~\ref{sec:grips:parametrize}) for covariance parameters with the goal of reducing posterior correlations and improve ESS of the resulting sampler. 

\section{Gridding for DAG-based models}\label{sec:grips:grid}
Gridding is usually understood as a form of discretization of a continuous spatial or temporal domain that leads to simplified structure and facilitates model fitting. For example, gridding may be used to fit spatial point-pattern models \citep{hmasd}, but it also arises as an intuitive idea with dynamic spatio-temporal models to describe the temporal dynamics of the process \citep{CressieWikle2011}. In this article, we use gridding as a modeling device: the ``grid'' is defined as a special set of spatial locations along which we model spatial dependence. We model the observed data as conditionally independent on the (unobserved) grid data. Because of the additional restrictions on spatial dependence that we impose at the grid locations via DAG-based methods, the resulting posterior sampling algorithms involve a substantial reduction in complexity, which ultimately leads to a reduction in compute time and an increase in overall efficiency. 

DAG-based methods for scaling GPs simplify the spatial dependence structure by assuming conditional independence of the outputs given neighboring outputs. These assumptions yield a sparse likelihood approximation to the full GP likelihood (often called Vecchia's approximation in the spatial literature); for any finite set $\calL$ of $n$ spatial locations, one obtains
\begin{align} \label{eq:wdensity_reference}
    p( \bw_{\calL} \given \btheta) \approx p^*(\bw_{\calL} \given \btheta) &= \prod_{\bl \in \calL} p( \bw(\bl) \given \bwpa{\bl}, \btheta),
\end{align}
where each observed location $\bl \in \calD$ corresponds to a node in the DAG, and $[\bl] \subset \calL$ is the set of locations mapped to parents of $\bl$. When $p(\cdot)$ is Gaussian, (\ref{eq:wdensity_reference}) is a product of multivariate Gaussian densities; $p^*(\bw_{\calL})$ is then Gaussian with a sparse precision matrix.
In order to define a standalone spatial process based on (\ref{eq:wdensity_reference}) over continuous spatial domains, one typically introduces a set $\calS \subset \calD$ of \textit{reference} locations. These locations are then mapped---individually or in groups---to nodes of a sparse DAG. The graphical model encodes assumptions of spatial conditional independence. 
All other locations (the \textit{non-reference} locations) may be assumed conditionally independent on the realization of the process at $\calS$. If $[\bl]$ is populated as a set of neighbors of $\bl$ in $\calS$, then one obtains the nearest-neighbor GP \citep[NNGP;][]{nngp}. Alternative models that share a similar setup include meshed Gaussian processes (MGPs; \citealt{meshedgp}), which are constructed by mapping the nodes of a pre-determined directed acyclic graph (DAG) to groups of locations constructed via domain tessellations or tiling, possibly recursively \citep{spamtrees}.

The above mentioned methods are flexible in the choice of $\calS$; in particular, MGPs may be set up for considerable additional computational gains when evaluating $p^*(\bw_{\calS}\mid \btheta)$ if $\calS$ is a grid, relative to other scalable DAG-based models. In fact, in an MGP one partitions the spatial domain, and thus $\calS$, into $M$ disjoint regions. Then, one has $p^*(\bw_{\calS}\mid \btheta) = \prod_{i=1}^M N(\bw_i; \bH_i \bw_{[i]}, \bR_i)$, where $\bH_i = \Cov_{i, [i]} \Cov_{[i]}^{-1}$ and $\bR_i = \Cov_i - \bH_i\Cov_{[i], i}$ must be computed for all $i=1,\dots,M$ at a cost that depends on the number of parents of each node $i$ in the DAG, which determines the size of $\Cov_{[i]}$. Assuming for simplicity that $n_{\calS} = M J$ and all DAG nodes have at most 2 parents as in Figure \ref{fig:qmgp_dag}, then the cost to compute $p^*(\bw_{\calS} \mid \btheta)$ is $O( n_{\calS} J^2)$. Using a Mat\'ern covariance and a gridded $\calS$, one can bring such cost to $O(1)$ because the number of $\bH_i$ and $\bR_i$ matrices that must be computed is constant with respect to $n_{\calS}$. Despite the substantial computational savings in choosing a gridded $\calS$, alternative choices for $\calS$ have only been explored in the context of low-rank GP methods, where the set $\calS$ is typically referred to as the set of ``knots'' or ``sensors,'' whose choice is not a trivial matter, see e.g. \cite{krauseetal2008} or \cite{pp_adaptive_knots}. In DAG-based GP models, it is common to pick $\calS$ as the set of observed locations $\calT$. 

To elucidate further, if we were to choose $\calS$ such that $\calT \cap \calS = \emptyset$, then we would infer about $\btheta$ by evaluating $p^*(\bw_\calT \given \btheta) = \int p^*(\bw_\calT \given \bw_\calS, \btheta) p^*(\bw_\calS\given \btheta) d\bw_\calS$. The assumed conditional independence and Gaussian base densities imply that $p^*(\bw_{\calS}\given \btheta) = N(\bw_{\calS}; \bzero, \tilde{\Cov}_{\calS})$ and $p^*(\bw_{\calT}\mid \bw_{\calS}, \btheta) = N(\bw_{\calT}; \bcH \bw_{\calS}, \bcR)$, where $\bcH$ is sparse lower-triangular or lower block-triangular and $\bcR$ is diagonal or block-diagonal, depending on the DAG. Marginally, $p^*(\bw_{\calT} \given \btheta) = N(\bw_{\calT}; \bzero , \bcH \tilde{\Cov}_{\calS} \bcH^\top + \bcR)$. There are no direct advantages associated to specific choices for $\calS$ when evaluating $p^*(\bw_{\calT}\given \btheta)$ because one needs to compute either $(\bcH \tilde{\Cov}_{\calS} \bcH^\top + \bcR)^{-1}$ or $(\tilde{\Cov}_{\calS}^{-1} + \bcH^{\top} \bcR^{-1} \bcH)^{-1}$, neither of which shares the same structure that a gridded $\calS$ might induce on $\tilde{\Cov}_{\calS}^{-1}$. As an alternative to this evaluation, one can proceed via MCMC by sampling $\bw_\calS$, then evaluating $p^*(\bw_\calT \given \bw_\calS, \btheta) p^*(\bw_\calS\given \btheta)$, like in the sequential and block samplers proposed in \cite{nngp} and the parallel samplers with graph coloring in \cite{meshedgp} and \cite{spamtrees}. However, sampling the latent process at $\calS$ as well as at non-reference locations $\calT \setminus \calS = \calU$ before updating $\btheta$ based on the target density $p^*(\bw_{\calU} \given \bw_{\calS}, \btheta) p^*(\bw_{\calS} \given \btheta)$ leads to inefficiencies due to the dimension of the latent variables. 

We can extend the advantages of using a gridded $\calS$ to scenarios in which locations in $\calT$ are irregularly spaced and $\calT \cap \calS = \emptyset$ by rewriting model (\ref{eq:regression}) hierarchically as:
\begin{equation} \label{eq:dag_model}
\begin{split}
    \by( \bl ) &= \bX(\bl)^\top \bbeta + \bw(\bl) + \beps(\bl)\;;\quad \beps(\bl) \sim N(\bzero, \bD)\;; \\
    \bw(\bl) &= \bH_{\bl} \bw_{[\bl]} + \bxi(\bl)\;;\quad \bxi(\bl) \sim N(\bzero, \bR_{\bl})\;,
\end{split}
\end{equation}
where $\bH_{\bl} = \Cov_{\bl, [\bl]} \Cov_{[\bl]}^{-1}$ and $\bR_{\bl} = \Cov_{\bl} - \Cov_{\bl, [\bl]} \Cov_{[\bl]}^{-1}\Cov_{[\bl], \bl}$ with $\Cov_{\bl} = \Cov(\bl, \bl)$, $\Cov_{[\bl]} = \Cov([\bl], [\bl])$. 
We can avoid sampling at non-reference locations by writing (\ref{eq:dag_model}) as
\begin{equation}\label{eq:gridmgp}
\by( \bl ) = \bX(\bl)^\top \bbeta + \bH_{\bl} \bw_{[\bl]} + \beps'(\bl)\;;\quad \beps'(\bl) \sim N(\bzero, \bD + \bR_{\bl}).
\end{equation}
We can interpret (\ref{eq:gridmgp}) as an extension of the bias-adjusted predictive process \citep[MPP;][]{modifiedpp}: the MPP and (\ref{eq:gridmgp}) coincide if the DAG includes a single reference node.
Because $\btheta$ indexes $\bH_{\bl}$ and $\bR_{\bl}$ for all $\bl \in \calT$, updating it via MCMC will now target the density $ N(\bw_{\calS}; \bzero, \tilde{\bC}_{\calS}) \prod_{\bl \in \calT} N(\by(\bl); \bX(\bl)^\top \bbeta, \bD + \bR_{\bl})$, where we exploit (\ref{sec:grips:grid}) and write $N(\bw_{\calS}; \bzero, \tilde{\bC}_{\calS}) = \prod_{\bl \in \calS} N(\bw(\bl); \bH_{\bl} \bwpa{\bl}, \bR_{\bl})$ to evaluate the process density at $\calS$ without building the sparse precision matrix $\tilde{\bC}^{-1}_{\calS}$. Writing the model as in (\ref{eq:gridmgp}) also allows us to take advantage of additional structure in $\calS$, as we outline in the next paragraph. The remaining portion of the density evaluation involves the product of $q$-dimensional densities, which is swiftly computed as we do not assume $q$ is large, and may be further simplified under an assumption of conditional independence of the outcomes given latent process realizations (i.e., if $\bD$ is diagonal).

\begin{figure}
  \centering
\begin{tikzpicture}
  \node[latent]  (w11) {$\bv_{11}$};
  \node[latent, right=of w11] (w12) {$\bv_{12}$};
  \node[latent, right=of w12] (w13) {$\bv_{13}$};
  \node[latent, above=of w11] (w21) {$\bv_{21}$};
  \node[latent, right=of w21] (w22) {$\bv_{22}$};
  \node[latent, right=of w22] (w23) {$\bv_{23}$};
  \node[latent, above=of w21] (w31) {$\bv_{31}$};
  \node[latent, right=of w31] (w32) {$\bv_{32}$};
  \node[latent, right=of w32] (w33) {$\bv_{33}$};
  \edge {w11} {w12} ; %
  \edge {w12} {w13} ; %
  \edge {w12} {w22} ; %
  \edge {w13} {w23} ; %
  \edge {w11} {w21} ; %
  \edge {w21} {w22} ; %
  \edge {w22} {w23} ; %
  \edge {w22} {w32} ; %
  \edge {w21} {w31} ; %
  \edge {w31} {w32} ; %
  \edge {w32} {w33} ; %
  \edge {w23} {w33} ; %
\end{tikzpicture}
  \caption{Reference nodes of a QMGP model DAG associated to a $3\times 3$ partition of $\calS$}
  \label{fig:qmgp_dag}
\end{figure}
Although the above discussion is general to any DAG-based scalable GP method, we now consider a ``cubic'' meshed GP \citep[QMGP;][]{meshedgp} in which a patterned DAG governs dependence across a gridded $\calS$, which may result in advantages in evaluating $p^*(\bw_{\calS} \mid \btheta)$---the largest time bottleneck among all MCMC steps. To build a QMGP, we create a sparse DAG by operating an axis-parallel tessellation of the spatial domain $\calD = [0, 1]^d$, obtaining $\calD = \cup_{i=1}^M \calD_i$. Then, we build a reference set $\calS$ and partition it into $M$ blocks $S_i$, $i=1, \dots, M$, where $S_i \subset \calD_i$. Block $S_i$ is mapped to the $i$th node in the DAG; the realization of $\bw$ at $S_i$ is $\bw_i = \bw(S_i) = (\bw(\bl_1)^\top, \dots, \bw(\bl_{n_{S_i}})^\top)^\top$. Similarly, $\bwpa{i}$ is the realization of $\bw$ corresponding to parents of the $i$th node. If the domain dimension is $d=2$, reference nodes in the DAG of a QMGP have at most two parents and two children as in Figure \ref{fig:qmgp_dag}. Our assumption here is that for any $i$, $S_i = S_1 + \delta$, where $\delta \in \Re^d$, meaning that each reference subset is the shifted version of the first. For example, a regular grid satisfies this assumption; we could thus let $\calS = \left\{ \left( \frac{j_1}{N_1},\dots, \frac{j_d}{N_d} \right) : (j_1, \dots, j_d) \in \{1, \dots, N_d\}^d \right\}$ so $n_S = \prod_{j=1}^d N_j$. We consider a grid with a different pattern in LiDAR data analysis of Section \ref{sec:applications:realdata}. Because $\calS$ has a shift-invariant grid pattern, evaluating $p^*(\bw_\calS \given \btheta)$ may be further sped up by choosing a shift-invariant cross-covariance function, in which case for all $S_i, i \in \{1,\dots,M\}$,  $p(\bw_{i} \given \bwpa{i}, \btheta) = N(\bw_i; \bH_j \bwpa{i}, \bR_j)$ for some $j$ such that $j \in \{i_1, \dots, i_{M^*} \} \subset \{1, \dots, M\}$, where $M^* \ll M$. Because $M^*$ is a small number that does not depend on the size of the dataset, the Gaussian conditional covariances are computed only $M^*$ times, and not $M \propto O(n)$ times, leading to massive computational speedups in evaluating $p^*(\bw_\calS \given \btheta)$. 
Writing the model as in (\ref{eq:gridmgp}) does not lead to major changes when sampling $\bw_\calS$, which proceeds as prescribed by the spatial DAG: for each node $i$ and corresponding subset $\calS_i$, we sample $\bw_i$ conditional on its Markov blanket. Additional details on the sampling algorithm with gridding are provided in Section \ref{sec:bayesian_grips_regression}. When at least one outcome is non-Gaussian likelihoods, we can use the same gridding method to infer about $\btheta$; its updates will proceed by targeting $N(\bw_{\calS}; \bzero, \tilde{\bC}_{\calS}) \prod_{\bl \in \calT} \prod_{j=1}^q p_j(\by(\bl) \given \bw_{[j]})$, where $p_j$ is the likelihood model for outcome $j$. The major computational bottleneck in a non-Gaussian setting is in sampling $\bw_{\calS}$. \cite{melange} introduces efficient non-marginalized samplers can be built via gradient-based methods that use second-order information about the target density.

\begin{figure}
\begin{subfigure}{.32\textwidth}
  \centering
  \begin{tikzpicture}
  \node[obs] (y) {$\by$};
  \node[latent, left=of y, yshift=.5cm] (w) {$\bw$};
  \node[const, above=of w] (empty) { };
  \node[latent, left=of w]  (theta) {$\btheta$};
  \node[latent, left=of y, yshift=-.5cm] (t) {$\bbeta, \btau$};
  \edge {w} {y} ; %
  \edge {t} {y} ; %
  \edge {theta} {w} ; %
\end{tikzpicture} 
\vspace{0.2cm}\\
\footnotesize \begin{tabular}{|c|}
\hline
    $\bw \sim N(\bzero, \Cov_{\btheta})$\\
    $\by = \bX\bbeta + \bw + \beps$\\ \hline
  \end{tabular}\normalsize
  \caption{Centered (\ref{eq:regression}) }
  \label{fig:param:centered}
\end{subfigure}
\begin{subfigure}{.32\textwidth}
  \centering
\begin{tikzpicture}
  \node[obs]                               (y) {$\by$};
  \node[latent, left=of y, yshift=1cm] (lam) {$\bLambda$};
  \node[latent, left=of y] (w) {$\bomega$};
  \node[latent, left=of w]  (phi) {$\bvarphi$};
  \node[latent, left=of y, yshift=-1cm] (t) {$\bbeta, \btau$};
  \edge {lam} {y} ; %
  \edge {w} {y} ; %
  \edge {t} {y} ; %
  \edge {phi} {w} ; %
\end{tikzpicture}
\vspace{0.2cm}\\
\footnotesize \begin{tabular}{|c|}
\hline
    $\bomega \sim N(\bzero, \brho_{\bvarphi})$\\
    $\by = \bX\bbeta + (\bI_n \otimes \bLambda)\bomega + \beps$\\ \hline
  \end{tabular}\normalsize
  \caption{Noncentered (\ref{eq:lmc_regression})}
  \label{fig:param:noncentered}
\end{subfigure}
\begin{subfigure}{.32\textwidth}
  \centering
\begin{tikzpicture}
  \node[obs]                               (y) {$\by$};
  \node[latent, left=of y, yshift=1cm] (lam) {$\bLambda$};
  \node[latent, left=of y, xshift=-0.5cm, yshift=0.2cm] (phi) {$\bvarphi$};
  \node[latent, left=of y, yshift=-0.7cm, xshift=-0.3cm]  (w) {$\bu$};
  \node[latent, left=of y, yshift=-1.2cm, xshift=0.6cm] (t) {$\bbeta, \btau$};
  \edge {lam} {y} ; %
  \edge {w} {y} ; %
  \edge {t} {y} ; %
  \edge {phi} {y} ; %
\end{tikzpicture}
\vspace{0.2cm}\\
\footnotesize \begin{tabular}{|c|}
\hline
    $\bu \sim N(\bzero, \bI_n)$, 
    $\bL_{\bvarphi}\bL_{\bvarphi}^\top = \brho_{\bvarphi}$\\
    $\by = \bX\bbeta + (\bI_n \otimes \bLambda)\bL_{\bvarphi} \bu + \beps$\\ \hline
  \end{tabular}\normalsize
  \caption{Alternative noncentered}
  \label{fig:param:altnoncentered}
\end{subfigure}
\caption{Some equivalent parametrizations of a spatial multivariate regression.}
\label{fig:parametrizations}
\end{figure} 

Finally, we provide some guidelines on grid design. Grid points should be located in the proximity of observed locations; if the data are located non-uniformly on the spatial domain, more grid points should be placed in areas with a higher density of observed locations---one example of this scenario is the application of Section \ref{sec:applications:realdata}. Otherwise, a regular grid suffices. 
The size of the grid and the partitioning scheme for fitting the QMGP should be determined by the available computational budget: a larger budget can afford a larger grid associated with a coarser partitioning. Our R package \texttt{meshed}, available on CRAN, provides the user with estimated runtimes that can be used in preliminary analyses to determine the size of the grid and the associated DAG.

\section{Parameter expansion for spatial big data}\label{sec:grips:parametrize}
For a constant $\sigma>0$, the random variables $u \sim N(0, \sigma^2)$ and $z=\sigma v$ where $v \sim N(0, 1)$ have the same distribution. Following the same principle, there are multiple parametrizations of a Bayesian hierarchical model (see, e.g., Figure \ref{fig:parametrizations}), leading to a multiplicity of posterior sampling algorithms with dissimilar efficiency profiles. Indeed, model parametrizations are known to impact performance of MCMC \citep{gelfandsahucarlin95, liuwu99, vandykmeng2001, yumeng11, kastnerfschnatter14, zanellaroberts21}, and there is an extensive literature on parameter expansion or related methods for improving algorithmic performance in particular models (\citealt{chuanhaietal1998, mengvandyk99, gelmancomment, gelman2004, lawrenceetal08, wangwest2009, ghoshdunson09, bhattacharya_dunson, rockovageorge16} and others). 
We are not aware of parameter expansion being used in geostatistical models, where the focus is the efficient estimation of the joint posterior of several parameters characterizing spatial covariance. Our primary goal is to use parameter expansion to improve efficiency in estimating such parameters.

The idea of using parameter expansion for algorithmic efficiency is appealing in Bayesian geostatistical models: high posterior dependence between $\sigma^2$ and $\phi$ and poor mixing of MCMC are common when dealing with large scale data at very high spatial resolution. As increasingly higher resolution sensors lead to larger quantities of data covering the same areas, we may explain high posterior dependence by the fixed-domain asymptotic properties of Mat\'ern covariances. In fact, only the microergodic parameter $\sigma^2 \phi^{2\nu}$ of the Mat\'ern model can be estimated consistently with an infinitely-sized sample in a fixed domain (\citealt{ying91} and Theorem 2 of \citealt{zhang04} and related corollaries; also refer to \citealt{stein90, nuggetconsist, vecchiaconsist}). This interpretation is more realistic than an increasing-domain setting in which an infinite-size sample is obtained by increasing the domain rather than by increasing resolution. Hence, it will be inappropriate to assume that $\sigma^2$ and $\phi$ are independent \textit{a priori}. Because our expansion method also implicitly induces a dependent prior for $\sigma^2$ and $\phi$, we also address the secondary goal of defining an appropriate dependent joint prior for $(\sigma^2,\phi)$. Choosing a prior distribution for these parameters is in general a difficult problem; see, e.g., related work in \cite{simpsonetal17} \cite{fuglstadetal19}.
We outline our proposed parameter expansion strategy below, first by illustrating our proposal in the simple univariate spatial regression case, then by considering a more general model for multivariate outcomes based on coregionalization or latent spatial factors using Mat\'ern margins.

\subsubsection{Univariate spatial regression on latent GP with exponential covariance}
Letting $q=k=1$, the centered parametrization (\ref{eq:regression}) becomes  $y(\bl) = \bx(\bl)^\top \bbeta + w(\bl) + \varepsilon(\bl)$, where $w(\cdot) \sim GP(\bzero, \sigma^2 \rho_{\nu}(\cdot, \cdot) )$ and $\rho_{\nu}(\cdot, \cdot)$ is the Mat\'ern covariance with smoothness $\nu$. We obtain a non-centered parametrization of the same model by letting $q=k=1$ in (\ref{eq:lmc_regression}), as $y(\bl) = \bx(\bl)^\top \bbeta + \lambda \omega(\bl) + \varepsilon(\bl)$, where $\omega(\cdot) \sim GP(\bzero, \rho_{\nu}(\cdot, \cdot) )$. In the latter model, $\lambda$ is only identifiable subject to a sign constraint (e.g., $\lambda >0$) but in practice we are interested in its square, since $\text{var}\{ \lambda \omega(\bl) \} = \lambda^2$. Although the two models share the same likelihood, choosing priors that are conditionally conjugate may lead to different posterior distributions for $\lambda^2$ and $\sigma^2$: the conditionally conjugate choice in the former is $1/\sigma^{2} \sim Gamma(\alpha_{\sigma}, 1/\beta_{\sigma})$, whereas in the second it is $\lambda \sim N(m_{\lambda}, v^2_{\lambda})$. 
Our proposal is to write the univariate regression model as 
\begin{equation} \label{eq:expanded_univariate}
\begin{aligned}
y(\bl) = \bx(\bl)^\top \bbeta + a r(\bl) + \varepsilon(\bl),
\end{aligned}
\end{equation}
where $r(\cdot) \sim GP(\bzero, s^2 \rho(\cdot, \cdot)/\phi^{2\nu} )$. The parameters $\sigma^2$ and $\lambda^2$ are identifiable in the former two models, whereas $s^2$ and $a^2$ are not identifiable in our overparametrization. However, the product $a^2 s^2 = \sigma^2 \phi^{2\nu}$ is identifiable; thus, by assigning priors to $a$ and $s^2$ we are implicitly defining a prior for $a^2 s^2$. In particular, we may seek conditionally conjugate updates for both $a$ and $s^2$ by letting $a$ have a Gaussian prior independent of an inverse gamma prior on $s^2$. These choices define an implicit half-t prior on the product $as$, as outlined in \cite{gelmancomment}. Conditionally conjugate priors do not involve rejection mechanisms in MCMC and are free of tuning requirements; they are thus likely more efficient than more general priors in these contexts. 
From the triplet $(a, s^2, \phi)$ we can revert to the identifiable parameters because $\text{var}\{ w(\bl) \} = \text{var}\{ \lambda \omega(\bl) \} = \sigma^2 = \lambda^2 = a^2 s^2 / \phi^{2\nu}$. In other terms, we can estimate identifiable model parameters using our overparametrized model.
Finally, we note that $a^2 s^2 = \sigma^2 \phi^{2\nu}$ is the Mat\'ern microergodic parameter. Because we implicitly assign a half-t prior on $as$, our expansion proposal also implicitly defines a dependent joint prior on $(\phi,\sigma^2)$, which as we mentioned is more appropriate than commonly used independent priors in the context of very high resolution data.

\subsubsection{Multivariate regression via Mat\'ern coregionalization and spatial latent factors}
For multivariate outcomes, we consider a LMC or factor model with Mat\'ern margins, as in (\ref{eq:lmc}) with $q\ge k>1$. The case $q=k$ is illuminating.
Since $\Cov_{\btheta}(\bzero) = \bLambda \brho(\bzero; \bvarphi) \bLambda^\top = \bLambda \bLambda^\top$ and $\brho(\bzero; \bvarphi) = \bI_k$ when $\bl - \bl' = 0$, we note that $\bLambda$ identifies with the lower-triangular Cholesky factor of $\Cov(\bzero)$ with positive diagonal entries \citep[see. e.g.,][and references therein for Bayesian LMC models]{finley2008,zhangbanerjee20}.  If $q>k$, one obtains a spatial factor model, where $\bLambda$ is identifiable subject to similar constraints (see, e.g., \citealt{Ren2013, taylor2019spatial, zhangbanerjee20}). Specifically, if $\bLambda$ is lower-triangular with positive diagonal entries, then $\bLambda = \bA \bSigma_{\lambda}$, where $\bA$ is lower-triangular with unit diagonal and $\bSigma_{\lambda} = \text{diag}\{\sigma_1, \dots, \sigma_k\}$ where $\sigma_j >0$ for $j=1,\dots,k$. Hence, analogously to the univariate case, we obtain two equivalent parametrizations for the multivariate process $\bw(\cdot)$:
\begin{equation}\label{eq:lmc_parametrizations}
\begin{matrix}
   \bw(\bl) = \bLambda \bomega(\bl) &\qquad \bw(\bl) = \bA \bv(\bl) \\
   \omega_j(\cdot) \sim GP(0, \rho_j(\cdot, \cdot)) &\qquad v_j(\cdot) \sim GP(0, \sigmasq_j \rho_j(\cdot, \cdot)),
\end{matrix}
\end{equation}
where the left expression coincides with (\ref{eq:lmc}).
We extend our proposed over-parametrization to the multivariate setting by writing the LMC or spatial factor model as
\begin{equation}\label{eq:lmc_expanded}
\begin{matrix}
   \bw(\bl) = \bcA \br(\bl)\\
   \br_j(\cdot) \sim GP(0, s^2_j \rho_j(\cdot, \cdot)/\phi_j^{2\nu}),
\end{matrix}
\end{equation}
where $\bcA$ is lower-triangular with positive diagonal entries and $\rho_j(\cdot, \cdot)$ is the Mat\'ern correlation function with smoothness $\nu$. 
%
Analogously to the univariate setting, we can use the overparametrized model (\ref{eq:lmc_expanded}) to estimate parameters of the identifiable parametrizations (\ref{eq:lmc_parametrizations}). For instance, by letting $\bJ = \text{diag}\{ \phi_1^{\nu}/s_1, \dots,  \phi_k^{\nu}/s_k\}$ we obtain $\bLambda = \bcA \bJ$.

In practice, we assign prior distributions to the elements of $\bcA$ -- a multivariate Gaussian on rows of $\bcA$ subject to sign constraints on the diagonal elements is analogous to the proposal in \cite{ghoshdunson09} for non-spatial models and yields conditionally conjugate updates with Gaussian data; analogously to the univariate case, $s_j^2, \phi_j$, for $j=1,\dots,k$ can be assigned inverse Gamma and uniform priors, respectively.

\section{Gridding and parameter expansion in Bayesian spatial regression}\label{sec:bayesian_grips_regression}
\begin{figure}[t]
    \centering
    \begin{tikzpicture}
  \node[obs]                               (y) {$\by$};
  \node[latent, left=of y, yshift=1cm] (lam) {$\bcA$};
  \node[latent, left=of y] (w) {$\br$};
  \node[latent, left=of w]  (phi) {$\bvarphi$};
  \node[latent, left=of y, yshift=-1cm] (t) {$\bbeta, \btau$};
  \node[const, right=of y, yshift=.8cm, xshift=1cm] (mean1) {$\bLambda = \bcA \bJ = \bcA \begin{smallbmat} \phi_1^{\nu}/s_1 & & \\ & \ddots & \\ & & \phi_k^{\nu}/s_k \end{smallbmat}$};
  \node[const, right=of y, yshift=-.8cm, xshift=1cm] (mean2) {$\bvarphi = \{ s^2_j, \phi_j \}_{j=1}^k$};
  \edge {lam} {y} ; %
  \edge {w} {y} ; %
  \edge {t} {y} ; %
  \edge {phi} {w} ; %
\end{tikzpicture}
\caption{DAG of the Bayesian spatial regression model using a \textit{PS} parametrization. The relationship with a standard parametrization is summarised on the right.}
\label{fig:grips_param}
\end{figure}

We combine gridding and parameter expansion (in short, \modelname) by parametrizing the LMC as in (\ref{eq:lmc_expanded}) within model (\ref{eq:gridmgp}), and obtain the resulting \modelname\ regression model: 
\begin{equation}\label{eq:gripsregression_location}
\begin{split}
\by( \bl ) &= \bX(\bl)^\top \bbeta + \bcZ_{\bl} \br_{[\bl]} + \beps'(\bl)\;;\quad \beps'(\bl) \sim N(\bzero, \bD + \bSigma_{\bl}),
\end{split}
\end{equation}
where $\br(\cdot) \sim QMGP(\bzero, \bK(\cdot))$ is a $k$-variate process with $\br(\bl) = (r_1(\bl), \dots, r_k(\bl))^{\top}$ and, for each $j=1, \dots, k$, $r_j(\bl) \sim QMGP\left(0, K_j(\cdot) \right)$ where $K_j(\cdot) = s^2_j \frac{\rho_j(\cdot)}{\phi_j^{2\nu}}$ with $\rho_j(\cdot)$ as the Matérn correlation in (\ref{eq:matern}), $\bcZ_{\bl} = \bcA \bH_{\bl}$, and $\bSigma_{\bl} = \bcA \bR_{\bl} \bcA^{\top}$. Compactly across locations $\calT$, we get
\begin{equation}\label{eq:gripsregression}
\begin{split}
\by &= \bX \bbeta + \bcZ \br + \beps' \;;\quad \beps' \sim N(\bzero, \bD_n + \bSigma)\;,
\end{split}
\end{equation}
where $\bcZ = (\bI_n \otimes \bcA) \bcH$, $\bSigma = (\bI_n \otimes \bcA)\bcR(\bI_n \otimes \bcA^\top)$, and $\otimes$ denotes the Kronecker product. In addition to the terms in (\ref{eq:regression}), we now explicitly have $\br$ representing the $n_{\calS} k \times 1$ vector of latent effects at the gridded reference set $\calS$. Furthermore, $\bcH$ is a sparse $nk \times n_{\calS}k$ matrix whose $i$th row block is filled with $\bH_{\bl_i}$, with zeroes at columns corresponding to locations in $\calS$ not in the parent set of $\bl_i$. Since the $k$ marginals for each process in $\br$ are independent, permuting the columns of $\bH_{\bl_i}$ that arrange it according to the process marginals is a block diagonal matrix whose $j$th (of the total $k$) row block is $K_j(\bl, [\bl]) K_j^{-1}([\bl])$. Similarly, $\bcR$ is a diagonal $nk \times nk$ matrix whose $i$th block is $\bR_{\bl_i}$, itself diagonal with $j$th diagonal element $K_j(\bl, \bl) - K_j(\bl, [\bl]) K_j^{-1}([\bl]) K_j([\bl], \bl)$. Also, $\bSigma$ is  $nq \times nq$ block diagonal with $q\times q$ blocks---the $i$th block is $\bcA \bR_{\bl_i} \bcA^{\top}$.

We complete the Bayesian model by specifying $\bbeta \sim N(\bm_{\beta}, \bV_{\beta})$ and $(\bcA)_{ij} \sim N(m_a, v_a)$, and independent zero-mean Gaussian distributions truncated below zero for the diagonal elements of $\bcA$. Large values for the prior variance for elements of $\bcA$ will reflect large prior uncertainty, although alternatives may be considered from the large literature on Bayesian latent factor models. We also assign independent Inverse Gamma priors for each $s^2_j$, $j=1, \dots, k$ and each $\tau^2_i$, $i=1, \dots, q$. As for the spatial decay parameters $\phi_j$, we let $\phi_j \sim U(l, u)$, where the values of $l>0$ and $u<\infty$ can be set upon considering prior knowledge on the outcomes' effective range. For instance, we may set $u$ to the value that leads to 0.05 spatial correlation (or any other desired value) at the minimum observed distance. Similarly, we may set $l$ to the value that leads to 0.95 spatial correlation at a pre-specified distance $d$. 
Figure \ref{fig:grips_param} summarises the \textit{PS} parametrization---\modelname\ regressions are noncentered on $\bcA$ and centered on $\bvarphi = \{s^2_j, \phi_j, \nu_j \}_{j=1}^k$. While MCMC methods for this model will sample from the posterior distribution of $\bcA$, $\bphi$, and $\br$, our ultimate goal is to use \modelname\ to estimate model (\ref{eq:lmc_regression}). We achieve this goal by using the fact that $\bLambda=\bcA \bJ$; samples of $s^2_j$, $j=1,\dots,k$ have no other purpose or interpretation other than the calculation of $\bLambda$. Next, we outline the sampling algorithms for estimation and prediction.

\subsection{Estimation and prediction} \label{sec:estimation}
Updating $\br$ follows the QMGP DAG. Denoting $T_i \subset \calT$ as the set of locations having $S_i$ as parent set, we express $\by_{T_i} = \bX_{T_i} \bbeta + \bcZ_{T_i} \br_i + \beps_{T_i}$ and $\beps_{T_i} \sim N(\bzero, \bD_{T_i} + \bSigma_{T_i})$. Let $\tilde{\by}_{T_i} = \bI_{T_i} (\by_{T_i} - \bX_{T_i} \bbeta)$ and $\tilde{\bD}_{T_i} = \bI_{T_i}(\bD_{T_i} + \bSigma_{T_i})\bI_{T_i}$, where $\bI_{T_i}$ is a diagonal matrix with zeros at elements corresponding to missing outcomes in $T_i$, ones elsewhere. If the $i$th node of the QMGP DAG is a parent of the $j$th node, then we can partition the columns of $\bH_j$ as $\bH_j = [\bH_i : \bH_{[j]\setminus i}]$, and similarly $\br_{[j]} = ( \br_{i}^\top, \br_{[j]\setminus i}^\top )^\top$, where $[j]\setminus i$ refers to the set of parents of $j$ excluding $i$. This yields a Gaussian full conditional distribution for $\br_i$ with mean $\bg_i$ and covariance $\bG_i$, where
\begin{align*}
    \bG_i^{-1} &= \bcZ_{T_i}^{\top} \tilde{\bD}_{T_i}^{-1} \bcZ_{T_i} + \bR_i^{-1} + \sum\limits_{j \in \ch{\bv_i}} \bH_{[j]\setminus i}^{\top} \bR_{j}^{-1}\bH_{[j]\setminus i}  \\
    \bG_i\bg_i &= \bR_i^{-1} \bH_i \br_{[i]} + \bcZ_{T_i}^{\top} \tilde{\bD}_{T_i}^{-1} \tilde{\by}_{T_i} + \sum\limits_{j \in \ch{\bv_i}} \bH_{[j]\setminus i}^{\top} \bR_{j}^{-1}\br_{[j]\setminus i}.
\end{align*}
Updating $\bbeta$ proceeds from its full conditional distribution, $N(\bbeta; \bm_{\bbeta}^{(n)}, \bV_{\bbeta}^{(n)})$, where $\bV_{\bbeta}^{^{(n)}-1} = \bV_{\bbeta}^{-1} + \bX^{\top} (\bD_n + \bSigma)^{-1} \bX$ and $\bV_{\bbeta}^{(n)} \bm_{\bbeta}^{(n)} = \bV_{\bbeta}^{-1}\bm_{\bbeta} + \bX^\top (\bD_n + \bSigma)^{-1} (\by - \bcZ \br)$. Updates for other unknowns are obtained from Metropolis steps; in our implementation, we use the robust adaptive Metropolis algorithm of \cite{vihola2012} to target an acceptance rate of about 0.23. For $\bcA$ and $\btau$, we target the conditional posterior density,
\begin{equation} 
 \pi(\bcA, \btau) \prod\limits_{i \in \calT} \left.N(\by_{\bl_i}; \bX_{\bl_i}\bbeta + \bcZ_{\bl_i} \br_{\bl_i}, \bD + \bSigma_{\bl_i})\right|_{\text{obs}},
\end{equation}
where $\pi(\bcA, \btau)$ is the joint prior distribution for $\{\bcA, \btau\}$, and ``obs'' restricts the product to observed data, i.e. each term is a $l$-dimensional Gaussian corresponding to the $l\leq q$ observed outcomes at spatial location $\bl_i$. For $\bvarphi$, the Metropolis update is applied to the conditional posterior density 
\begin{equation}
    \pi(\bvarphi) \prod_{i}^M N(\br_i; \bH_{i} \br_{[i]}, \bR_{i}) \prod\limits_{i \in \calT} \left.N(\by_{\bl_i}; \bX_{\bl_i}\bbeta + \bcZ_{\bl_i} \br_{\bl_i}, \bD + \bSigma_{\bl_i})\right|_{\text{obs}},
\end{equation} 
where $\pi(\bvarphi)$ is the prior distribution for $\bvarphi$. It is computationally easier to update $\bcA$, $\btau$ and $\bolds{s}^2$ than $\bphi$ as the latter requires recomputing the correlations. 
An account of the computational time complexity of QMGP with \modelname\ is available in the Supplement. 
	
\section{Applications on synthetic data} \label{sec:applications:simulations:nongridded}
Our main goal is to assess the efficiency of \modelname\ in estimating covariance hyperparameters in univariate and multivariate settings. We validate our results using predictive performance. Our baseline for comparison includes a QMGP model without \modelname, i.e., using the observed data locations as the reference set, and without parameter expansion. In addition to \modelname, we also consider QMGP models in which we only use a gridded reference set without parameter expansion (\qmgp-Gri), and QMGP models in which we only use parameter expansion, but fix the reference set at the observed locations (\qmgp-PS).
When possible, we compare our proposed methods to alternative MCMC schemes for QMGP models, to different models (i.e. NNGP); and to an MCMC-free method such as INLA. All methods are allowed to exploit up to 12 CPU threads in a workstation with 128GB memory and AMD Ryzen 9 5950X CPU. In all our analyses, we allow each method to operate on a time budget of about 30 seconds.

\subsection{Univariate outcomes} \label{sec:sim:univariate}
Each univariate-outcome data set includes a realization of a Gaussian process with Mat\'ern covariance (\ref{eq:matern}) at $n=10000$ irregularly-spaced locations in $[0,1]^2$, different for each data set. The process smoothness $\nu$ is fixed to its known value (either $\nu=0.5$ or $\nu=1.5$). For all data sets, we use the same regular grid of $n_{\text{out}} = 10000$ spatial locations for out-of-sample predictions. For $\nu = 0.5$, we generate 50 datasets for each combination of $\sigmasq \in \{ 1, 5\}$ and $\phi \in \{1, 5, 16\}$, and similarly for $\nu = 1.5$ but with $\phi \in \{ 2, 6.5, 25 \}$, for a total of 600 datasets, each of size $n_{\text{all}} = 20000$. The effective spatial range $R_{\text{eff}}$ in our simulated datasets--i.e., the distance at which covariance equals 0.05--is shortest, $R_{\text{eff}} = 0.1872$, using $\sigma^2=1, \nu=0.5, \phi=16$; longest, $R_{\text{eff}} = 3.319$, using $ \nu=1.5, \sigma^2=5, \phi=2$. An example in which the longest range scenario would be realistic is a setting in which data with an effective range of $1km$ is imaged using a 36 million pixel sensor covering an area of 2km$\times$2km, but a smaller dataset of 1 million observations covering 300m$\times$300m is used for analysis.

To showcase the improvements from \modelname\ within MCMC schemes, which require posterior sampling of the latent process, we compare \qmgp-\modelname\ to the original Gibbs sampler of \cite{meshedgp} (\qmgp-Gibbs) which updates $\sigma^2$ using its full conditional distribution rather than in a block with $\phi$, and the analogous Gibbs sampler for NNGP models (\nngp-Gibbs, also referred to as the ``latent'' or sequential sampler, see \citealt{nngp_algos}). Furthermore, we offer comparisons with an NNGP model of the response \cite{zbf21}. Since it involves no sampling of the latent random effects (\nngp-Response), it is expected to produce more efficient sampling of covariance hyperparameters. We also consider a MCMC-free method based on stochastic partial differential equations \citep{spde} fitted via INLA (\textit{SPDE-INLA}). All QMGP models were setup on a $20\times 20$ axis-parallel tessellation of the spatial domain, whereas the NNGP models used $m=9$ neighbors, and \textsc{SPDE-INLA} was built with a mesh of $90\times 90$ points. MCMC chains for QMGP and NNGP methods were of length $5000$; of these, $2500$ iterations were removed as burn-in. Our setup ensured that the \qmgp-GriPS, \nngp-Response, and \textsc{SPDE-INLA} were computed with comparable time budgets on average (29.5, 32.4, 24.1 seconds, respectively, on our workstation). The combined runtime for all methods was about 70 hours. 

\begin{table}[]
    \centering
\resizebox{.75\columnwidth}{!}{%
\begin{tabular}{|l|ll|rrr|rrr|}
  \multicolumn{9}{l}{\textit{Efficiency}}\\
\hline
\multirow{2}{*}{Model} & \multirow{2}{*}{\begin{tabular}{l}Sampling\\method\end{tabular}} & & \multicolumn{3}{c|}{$\nu=0.5$} & \multicolumn{3}{c|}{$\nu=1.5$}\\
 &  &  & ESS & ESS/s & Relative & ESS & ESS/s & Relative \\ 
  \hline
\multirow{15}{*}{ QMGP} & \multirow{3}{*}{GriPS} & $\phi$ & 233.87 & 7.97 & \textbf{17.96} & 133.30 & 4.38 & \textbf{54.47} \\ 
   &  & $\sigmasq$ & 219.89 & 7.39 & \textbf{15.19} & 230.84 & 7.60 & \textbf{74.62} \\ 
   &  & $\tau^2$ & 93.07 & 3.18 & 0.68 & 335.99 & 11.12 & 6.70 \\ 
   \cline{2-9} & \multirow{3}{*}{PS} & $\phi$ & 259.94 & 4.60 & 10.36 & 133.06 & 2.29 & 28.42 \\ 
   & & $\sigmasq$ & 257.54 & 4.54 & 9.33 & 234.64 & 4.03 & 39.53 \\ 
   & & $\tau^2$ & 116.00 & 2.03 & 0.43 & 903.02 & 15.78 & 9.50 \\ 
   \cline{2-9}
 & \multirow{3}{*}{Gri} & $\phi$ & 193.08 & 7.04 & 15.86 & 103.60 & 3.58 & 44.53 \\ 
   &  & $\sigmasq$ & 127.96 & 4.52 & 9.30 & 172.41 & 5.97 & 58.59 \\ 
   &  & $\tau^2$ & 112.98 & 4.05 & 0.86 & 313.67 & 10.93 & 6.59 \\
  \cline{2-9} & \multirow{3}{*}{baseline} & $\phi$ & 254.14 & 4.51 & 10.15 & 72.81 & 1.27 & 15.81 \\ 
   &  & $\sigmasq$ & 227.69 & 3.97 & 8.15 & 151.82 & 2.64 & 25.92 \\ 
   &  & $\tau^2$ & 110.79 & 1.97 & 0.42 & 879.15 & 16.07 & 9.68 \\ 
 \cline{2-9}   & \multirow{3}{*}{Gibbs} & $\phi$ & 5.97 & 0.18 & 0.41 & 5.03 & 0.15 & 1.81 \\ 
   &  & $\sigmasq$ & 7.22 & 0.22 & 0.45 & 8.07 & 0.23 & 2.30 \\ 
   &  & $\tau^2$ & 137.20 & 4.19 & 0.89 & 878.33 & 25.52 & \textbf{15.37} \\ 
   \hline
\multirow{6}{*}{ NNGP}  & \multirow{3}{*}{Response} & $\phi$ & 14.37 & 0.44 & 1.00 & 12.89 & 0.08 & 1.00 \\ 
   &  & $\sigmasq$ & 15.75 & 0.49 & 1.00 & 16.14 & 0.10 & 1.00 \\ 
   &  & $\tau^2$ & 151.53 & 4.68 & \textbf{1.00} & 260.96 & 1.66 & 1.00 \\ 
 \cline{2-9}
   & \multirow{3}{*}{Gibbs} & $\phi$ & 6.57 & 0.16 & 0.35 & 3.21 & 0.02 & 0.23 \\ 
   &  & $\sigmasq$ & 9.17 & 0.22 & 0.45 & 8.47 & 0.05 & 0.49 \\ 
   &  & $\tau^2$ & 163.52 & 3.87 & 0.83 & 1195.78 & 7.02 & 4.23 \\ 
   \hline
\end{tabular}
}
    \caption{MCMC efficiency of \modelnames\ in the estimation of Matérn parameters in univariate settings. For posterior samples of size 2500, we report effective sample size---absolute, per unit time, and relative to the \nngp-Response model.}
    \label{tab:estimation_efficiency}
\end{table}

\begin{figure}
    \centering
    \includegraphics[width=\textwidth]{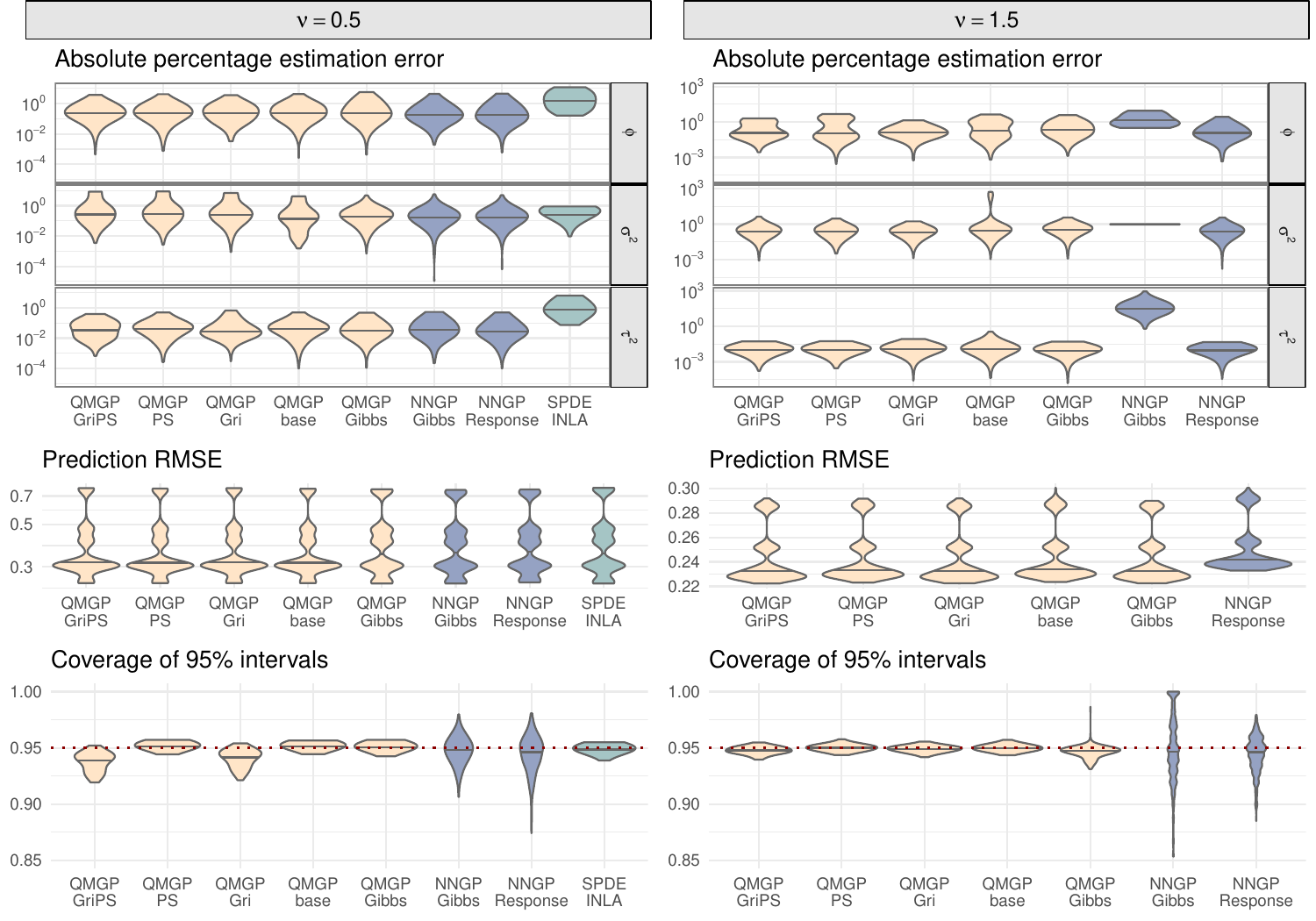}
    \caption{Estimation and prediction performance across the 600 univariate datasets of Section \ref{sec:sim:univariate}. Top row: violin plots of absolute percentage errors in estimating unknown model parameters; center row: root mean square errors (RMSE) in out-of-sample predictions; bottom row: empirical coverage of 95\% prediction intervals. Setting $\nu=1.5$ is unsupported in \texttt{R-INLA} and generated errors in predictions from the latent model in \texttt{spNNGP}}
    \label{fig:uni:estim_predict_covg}
\end{figure}

We measure sampling efficiency using effective sample size per unit time (ESS/s), reported in Table~\ref{tab:estimation_efficiency}. As expected, \qmgp-Gibbs and \nngp-Gibbs perform similarly, both less efficient than \nngp-Response with $\nu=0.5$. When $\nu=1.5$, \qmgp-Gibbs is more efficient than both \nngp methods. Relative to \nngp-Response, we may attribute the improved efficiency of the baseline QMGP method to the robust adaptive Metropolis algorithm \citep{vihola2012}. Our proposed approach leads to additional efficiency gains: relative to the baseline QMGP, PS leads to improved ESS in the $\nu=1.5$ setting. We finally notice that the inclusion of a reference grid (Gri) has a small negative impact on ESS; however, the substantial computational speedup more than makes up for the loss in ESS, resulting in much higher ESS per unit time. We achieve these improvements while maintaining excellent inferential performance, as seed in Figure \ref{fig:uni:estim_predict_covg}.

\subsection{Multivariate outcomes}\label{sec:sim:multivariate}

\begin{table}[]
\centering
\resizebox{.9\columnwidth}{!}{
\begin{tabular}{|l|rrr|rrr|rrr|rrr|}
  \cline{2-13}
\multicolumn{1}{c|}{} & \multicolumn{3}{c|}{baseline} & \multicolumn{3}{c|}{PS} & \multicolumn{3}{c|}{Gri} & \multicolumn{3}{c|}{GriPS} \\ 
\multicolumn{1}{c|}{} & ESS & ESS/s & Relative  & ESS & ESS/s & Relative & ESS & ESS/s & Relative  & ESS & ESS/s & Relative \\
  \hline
  $\Lambda_{11}$ & 4.22 & 0.05 & 1.00 & 121.60 & 1.51 & 28.88 & 3.30 & 0.07 & 1.30 & 107.27 & 2.19 & \textbf{41.82} \\ 
  $\Lambda_{21}$ & 11.38 & 0.14 & 1.00 & 62.76 & 0.78 & \textbf{5.51} & 5.32 & 0.11 & 0.77 & 31.87 & 0.65 & 4.60 \\ 
  $\Lambda_{22}$ & 5.98 & 0.07 & 1.00 & 112.10 & 1.39 & 18.77 & 5.73 & 0.12 & 1.58 & 120.94 & 2.46 & \textbf{33.22} \\ 
  $\Lambda_{31}$ & 8.71 & 0.11 & 1.00 & 52.40 & 0.65 & \textbf{6.03} & 4.46 & 0.09 & 0.85 & 22.36 & 0.46 & 4.22 \\ 
  $\Lambda_{32}$ & 7.94 & 0.10 & 1.00 & 47.12 & 0.58 & \textbf{5.94} & 5.52 & 0.11 & 1.15 & 21.17 & 0.43 & 4.38 \\ 
  $\Lambda_{33}$ & 8.16 & 0.10 & 1.00 & 116.73 & 1.45 & 14.31 & 3.53 & 0.07 & 0.72 & 80.73 & 1.64 & \textbf{16.23} \\ 
  \hline
  $\phi_1$ & 4.24 & 0.05 & 1.00 & 80.46 & 1.00 & 19.01 & 8.98 & 0.18 & 3.51 & 85.05 & 1.73 & \textbf{32.98} \\ 
  $\phi_2$ & 8.13 & 0.10 & 1.00 & 78.05 & 0.97 & 9.61 & 8.68 & 0.18 & 1.77 & 92.56 & 1.88 & \textbf{18.69} \\ 
  $\phi_3$ & 11.05 & 0.14 & 1.00 & 91.19 & 1.13 & 8.25 & 9.14 & 0.19 & 1.37 & 75.48 & 1.54 & \textbf{11.21} \\ 
  \hline
  $\tau^2_1$ & 45.26 & 0.56 & 1.00 & 77.80 & 0.97 & \textbf{1.72} & 29.43 & 0.60 & 1.08 & 44.62 & 0.91 & 1.62 \\ 
  $\tau^2_2$  & 53.53 & 0.66 & 1.00 & 61.33 & 0.76 & \textbf{1.15} & 26.68 & 0.55 & 0.82 & 29.51 & 0.60 & 0.91 \\ 
  $\tau^2_3$ & 51.50 & 0.64 & 1.00 & 60.61 & 0.75 & 1.18 & 32.08 & 0.66 & 1.03 & 55.91 & 1.14 & \textbf{1.78} \\ 
   \hline
\end{tabular}
}
\caption{MCMC efficiency of \modelname\ in estimating $\bLambda, \phi_j, \tau^2_j$, $j=1,2,3$ in multivariate settings.}\label{table:multivariate:efficiency}
\end{table}
\begin{table}[]
    \centering
    \resizebox{.98\columnwidth}{!}{
    \begin{tabular}{|l|l|rrrrrr|rrrrrr|rrr|rrr|}
  \hline
Model & Method & $B_{11}$ & $B_{12}$ & $B_{13}$ & $B_{21}$ & $B_{22}$ & $B_{23}$ & $\Lambda_{11}$ & $\Lambda_{21}$ & $\Lambda_{22}$ & $\Lambda_{31}$ & $\Lambda_{32}$ & $\Lambda_{33}$ & $\phi_1$ & $\phi_2$ & $\phi_3$ & $\tau^2_1$ & $\tau^2_2$  & $\tau^2_3$ \\ 
  \hline
\multirow{4}{*}{QMGP} &
baseline & \textbf{0.95} & 0.94 & 0.95 & 0.97 & 0.96 & \textbf{0.95} & 0.86 & 0.85 & 0.45 & 0.85 & 0.54 & 0.63 & 0.88 & 0.38 & \textbf{0.85} & \textbf{0.94} & 0.18 & 0.00 \\ 
  & Gri & 0.96 & 0.92 & 0.93 & \textbf{0.96} & 0.96 & 0.94 & 0.75 & 0.75 & 0.35 & 0.80 & 0.44 & 0.61 & 0.49 & 0.50 & 0.58 & 0.73 & 0.39 & 0.99 \\ 
& PS & \textbf{0.95} & \textbf{0.95} & \textbf{0.95} & \textbf{0.96} & 0.96 & \textbf{0.95} & \textbf{0.93} & \textbf{0.91} & \textbf{0.93} & \textbf{0.88} & \textbf{0.93} & \textbf{0.93} & \textbf{0.93} & 0.83 & 0.68 & 0.91 & 0.29 & 0.00 \\ 
& GriPS & 0.96 & 0.93 & 0.92 & \textbf{0.96} & \textbf{0.95} & 0.93 & 0.92 & 0.82 & 0.93 & 0.70 & 0.66 & 0.83 & 0.64 & 0.83 & 0.74 & 0.77 & \textbf{0.59} & \textbf{0.96} \\ 
\hline
  SPDE &
INLA & \textbf{0.95} & 0.96 & \textbf{0.95} & \textbf{0.96} & \textbf{0.95} & 0.93 & 0.09 & 0.60 & 0.13 & 0.65 & 0.70 & 0.14 & 0.85 & \textbf{0.85} & 0.81 & 0.00 & 0.00 & 0.00 \\ 
   \hline
\end{tabular}
}
    \caption{Empirical coverage of 95\% posterior intervals for unknown model parameters.}
    \label{table:multivariate:covg}
\end{table}

\begin{figure}
    \centering
    \includegraphics[width=\textwidth]{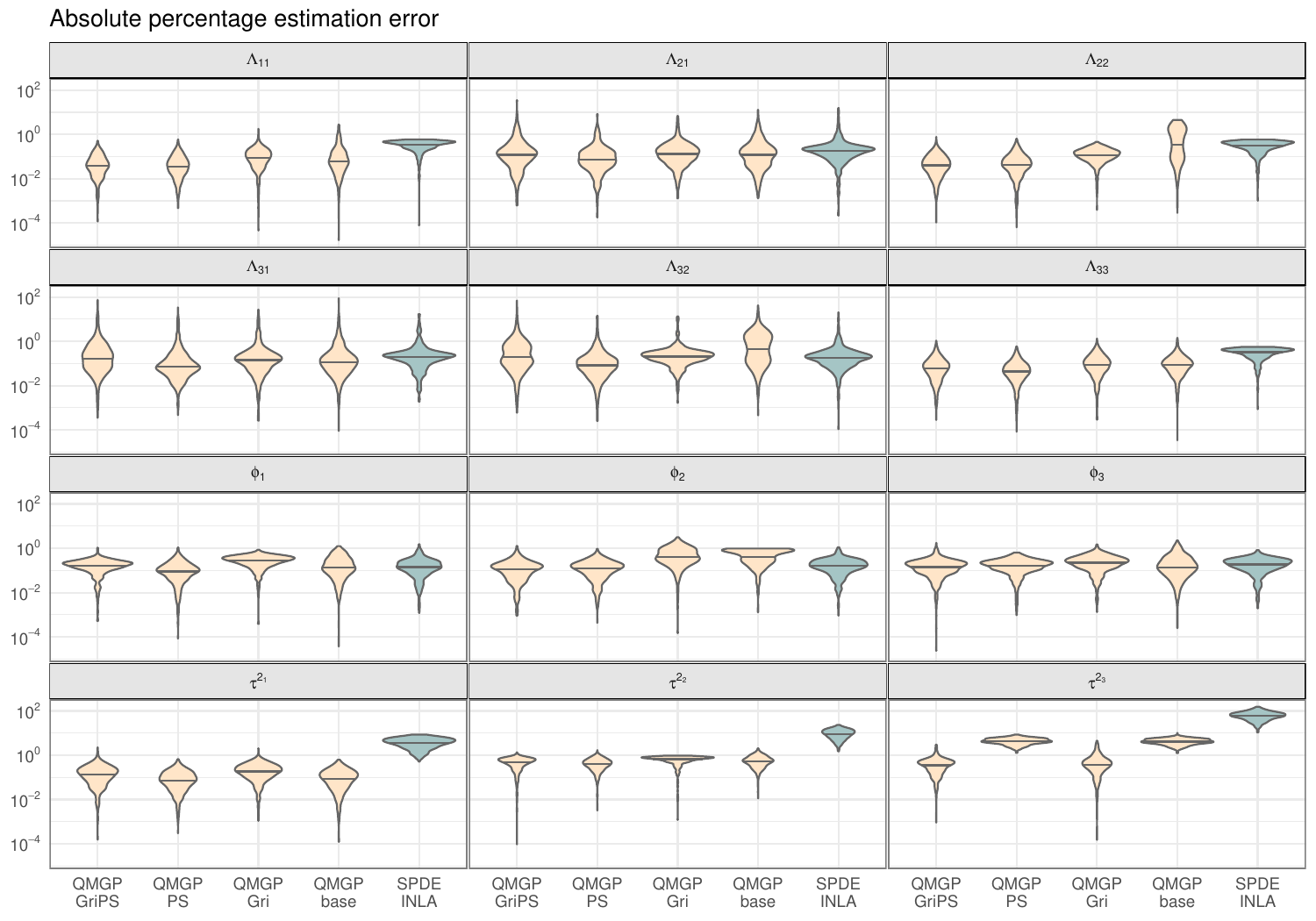}
    \includegraphics[width=\textwidth]{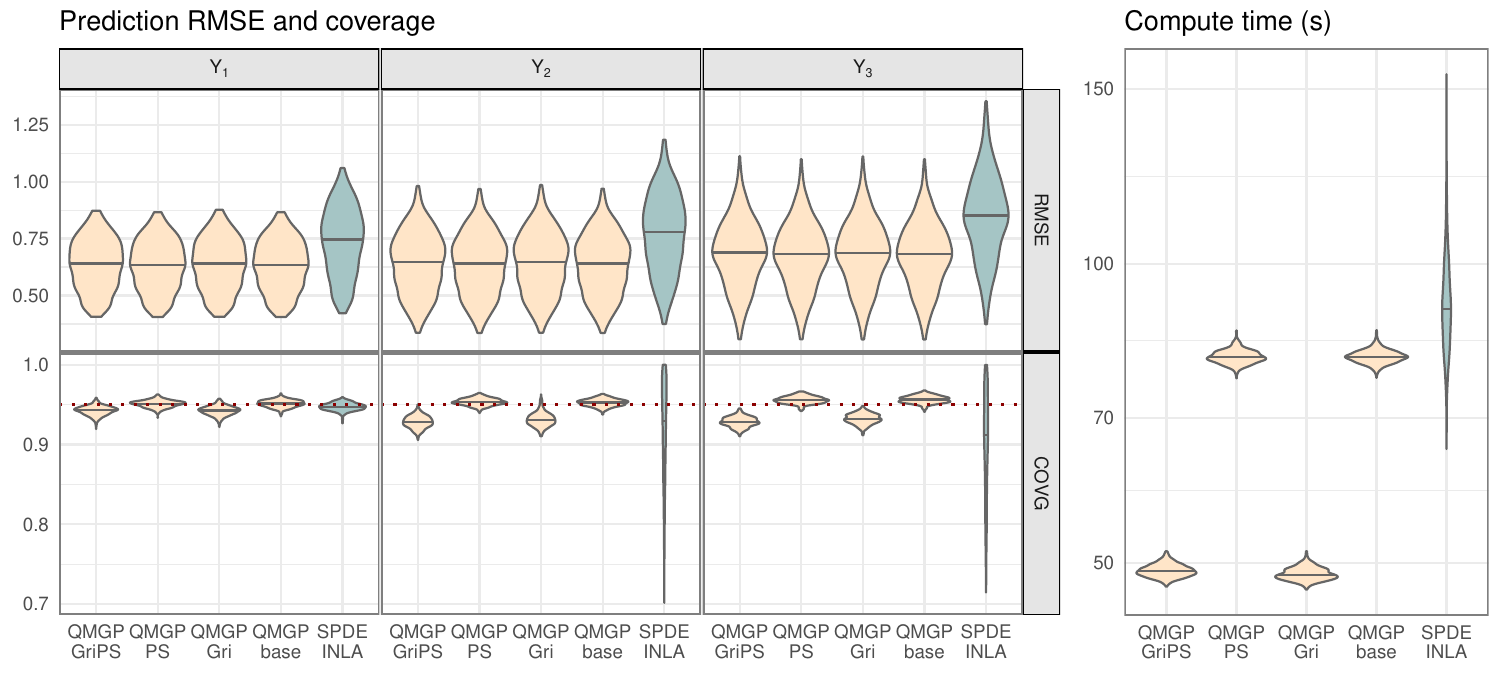}
    \caption{Estimation and prediction performance across the 500 multivariate datasets of Section \ref{sec:sim:multivariate}. Top: violin plots of absolute percentage errors in estimating unknown model parameters; Bottom-left: root mean square errors (RMSE) in out-of-sample predictions, and the corresponding empirical coverage of 95\% intervals. Bottom-right: compute time in seconds.}
    \label{fig:mv:performance}
\end{figure}

We simulate 500 data sets from model (\ref{eq:lmc_regression}) with $q=3$ outcomes and $p=2$ covariates. For each data set, we sample a different set of locations $S$ (of size $n_S = 4900$) uniformly in $[0,1]^2$; then, each outcome is fixed as missing at $n_i/2 = 2450$ locations selected uniformly at random -- this implies that for location $s \in S$, the probability that 0, 1, 2, or 3 outcomes are observed at $s$ are found as $\frac{1}{8}, \frac{3}{8}, \frac{3}{8}, \frac{1}{8}$, respectively and therefore, at least one outcome is missing at 87.5\% of spatial locations on average, and at least one is observed with the same proportion. This also implies that the effective dimension is $n_e = 7350$.
For each of the 500 datasets, we fix $(\tau_1^2, \tau_2^2, \tau_3^2)^\top = (0.1, 0.05, 0.01)^\top$ and $(\phi_1, \phi_2, \phi_3)^\top = (1, 6, 15)^\top$. We then generate elements of $\bB$ as $B_{ij} \sim N(0,1)$ independently, and finally construct the lower-triangular matrix $\bLambda = \bL \bD^{\frac{1}{2}}$ where $\bL$ is a lower-triangular matrix whose off-diagonal elements are $l_{ij} \sim U(-1, 1)$ and with unitary diagonal, whereas $\bD = (d_1, d_2, d_3)$ where $d_j \sim U(1,2)$. This implies that $\bw(s) = \bLambda \bomega(s)$ is a GP with cross-covariance $\Cov(\cdot; \btheta)$ such that $\Cov(\bzero; \btheta)=\bLambda \bLambda^\top = \bL \bD \bL^\top$.
We analyze the efficiency of QMGP models using combinations of Gri and PS, and compare their performance to \textsc{SPDE-INLA} in estimating $\bLambda, \tau^2_j$ and $\phi_j$ for $j=1,2,3$, and in predicting each of the outcomes on a grid of $n_{\text{out}} = 4900$ fully unobserved spatial locations. Figure \ref{fig:mv:performance} summarizes the result across all data sets. Table \ref{table:multivariate:efficiency} reports the efficiency figures. We observe that PS alone induces major efficiency gains compared to QMGP models without PS. While reference grids seemingly lead to a slightly worse coverage of posterior intervals in our simulated settings (bottom left of Figure \ref{fig:mv:performance} and Table \ref{table:multivariate:covg}), they also correspond to a 50\% reduction in compute times as shown in the bottom-right of Figure \ref{fig:mv:performance}, improved efficiency (Table \ref{table:multivariate:efficiency}), and similar prediction performance. The poor coverage of the \textsc{SPDE-INLA} method with regard to the nuggets is likely due to the number of mesh grid points (400) used in its implementation; with a relatively small number of grid points, the resulting approximation may overestimate the nugget term. However, increasing the number of grid points leads to much increased compute time in multivariate settings, exceeding the time budget allocation we reserved for fair comparisons with other models.

\section{Alaskan LiDAR data} \label{sec:applications:realdata}

\begin{figure}
    \centering 
        \includegraphics[width=0.45\textwidth]{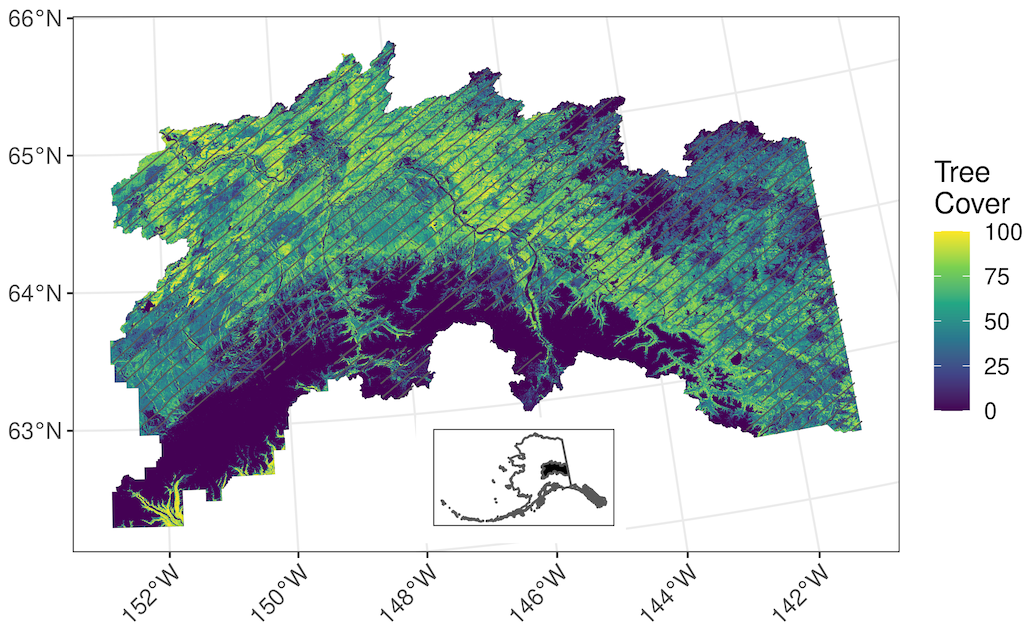}
        \includegraphics[width=0.45\textwidth]{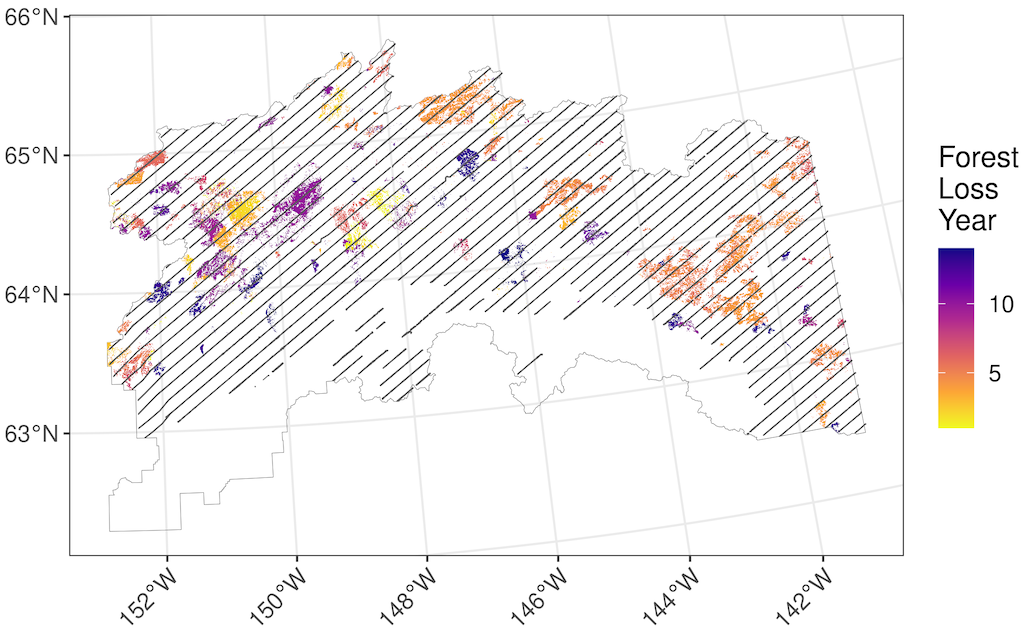}
        \includegraphics[width=0.45\textwidth]{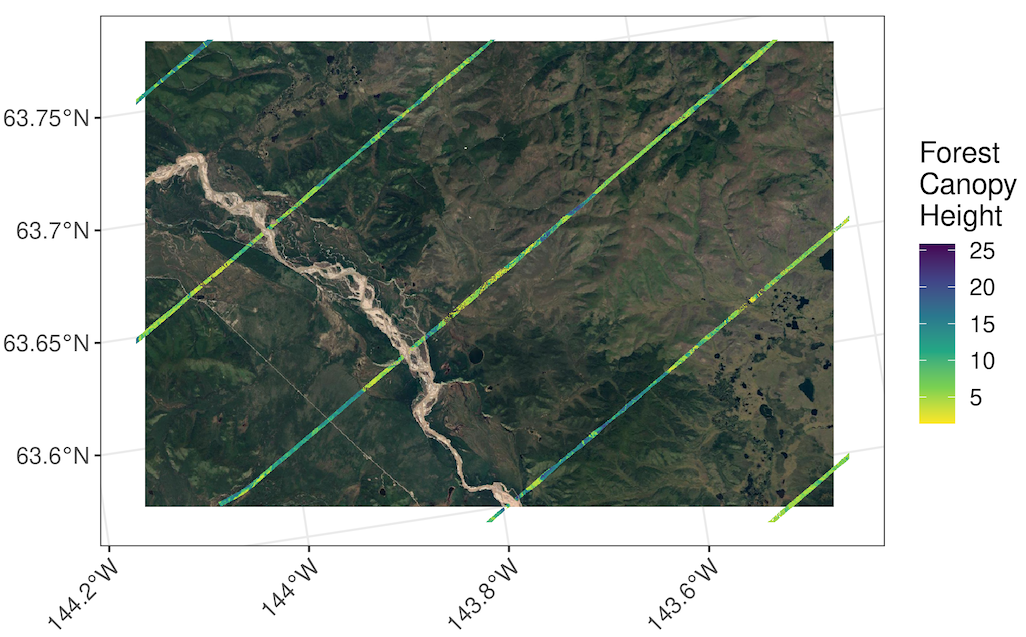}
        \includegraphics[width=0.45\textwidth]{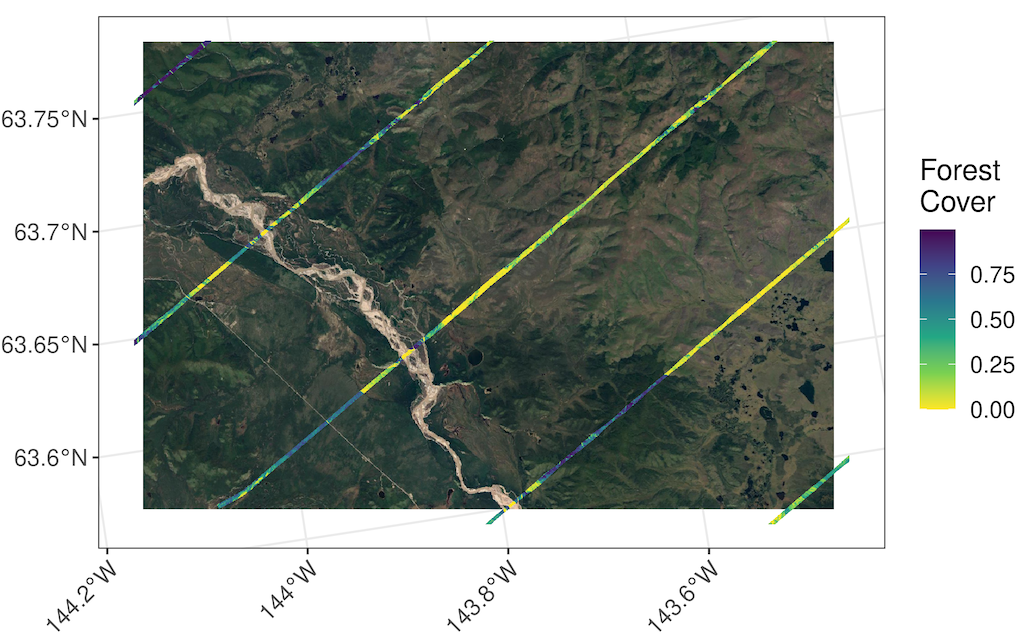}
    \caption{Top: tree cover in 2000 (left) and forest loss year (right) over the entire study area. Bottom: the two outcomes (\textit{p.90} and \textit{f.cvr}) shown at a smaller subregion. The strips along which the outcomes are recorded appear as thin lines in the top maps.}\label{fig:predictors_and_outcomes}
\end{figure}
We consider a large data set consisting of remotely sensed light detection and ranging (LiDAR) measurements collected using the National Aeronautics and Space Administration (NASA) Goddard's LiDAR, Hyperspectral, and Thermal (G-LiHT) Airborne Imager \citep{cook2013} sensor in Summer 2014 over a large region in Alaska. The data set includes observations of forest canopy height (labeled \textit{p.90} and measured in meters) and fractional forest cover (\textit{f.cvr}) at 12,270,334 spatial locations, which we seek to jointly characterize in a multivariate regression model. Owing to the low flying altitude (355m) and narrow field of view (30\degree) of the sensor, the observed locations are not uniformly spread out in the study area. Instead, they are highly concentrated along swaths less than 200m wide and separated by 9km of unobserved areas. This data has been analyzed earlier in univariate spatial regression modeling contexts \citep[see][for additional details]{nngp_algos}. 

We select a sub-region of the data of size $150\text{km} \times 250\text{km}$ covering about 31.2\% of the observed data, and sample 2.5 million spatial locations uniformly at random in this area to build the data set we consider. The resulting study area includes 17 strips of observed data. The effective dimension of our data is thus $nq = 5 \times 10^6$ accounting for $q=2$. We seek predictions for both outcomes on the full sub-region (including the wide areas separating the observed strips) and estimate the standard LMC spatial regression model (\ref{eq:lmc_regression}). We approach this problem by estimating the GriPS regression model (\ref{eq:gripsregression}) and subsequently recovering $\bLambda$ from $\bcA$ and $\sigmasq_j$ as detailed above. 

We consider two covariates that are available across the entire study area as shown in Figure~\ref{fig:predictors_and_outcomes}: (i) percent forest cover measured in the year 2000 (labeled \textit{tc}); and (ii) an indicator variable of forest loss (\textit{loss}) that is used to obtain the number of years since 2000 that a location had a forest-destroying disturbance such as a fire. Outcomes and predictors are shown in Figure~\ref{fig:predictors_and_outcomes}.

\begin{figure}
    \centering
    \includegraphics[width=\textwidth, trim=0 100 0 0, clip]{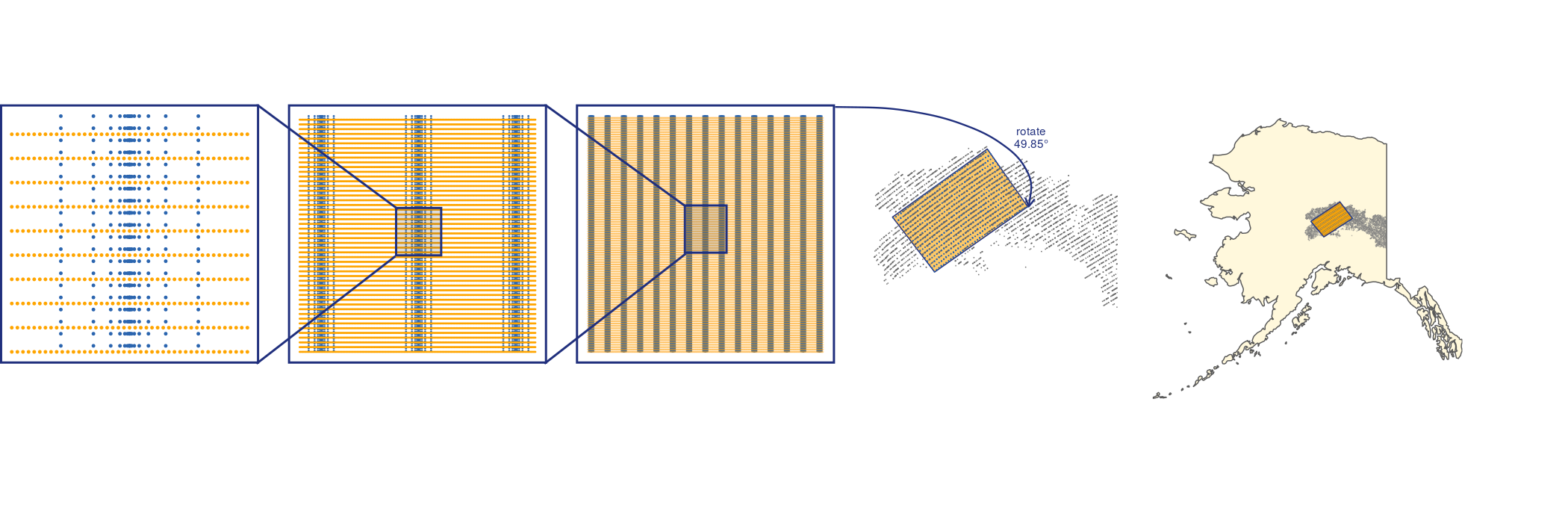}
    \caption{Construction of $\calS$ by overlaying a grid of regularly-spaced locations (orange) to a patterned grid (blue); alignment with the data swaths is achieved by rotating $\calS$.}
    \label{fig:refset_construction}
\end{figure}

When building a gridded reference set $\calS$ for this analysis, we aim to (1) match the data collection pattern in swaths, (2) retain a pattern for computational efficiency using QMGP. We thus construct $\calS$ as the union of two reference subsets, which we label $\calS_{\text{orange}}$ and $\calS_{\text{blue}}$. The former is a grid of regularly-spaced locations of size $n_{\text{orange}} = 1.52 \times 10^6$ which we use to uniformly cover the selected study area. Because only a small portion of locations in $\calS_{\text{orange}}$ are near the observed outcomes, using $\calS_{\text{orange}}$ would result in oversmoothing the latent surface, much like a low-rank GP model \citep[see, e.g.,][]{sudipto_ba17}. Therefore, we build an additional patterned grid of size $n_{\text{blue}} = 5.44\times 10^5$ to cover the 17 observed swaths.  
The construction of the two grids is visualized in Figure~\ref{fig:refset_construction}.
We construct a QMGP model on $\calS$ by partitioning the coordinates into $33$ and $500$ intervals, respectively. This partitioning scheme reflects the arrangement of the measurement locations: the number of intervals along the horizontal axis matches the number of data strips, resulting in better caching of matrix computations within QMGP-GriPS and thus in lower run times. The partitioning along the vertical axis was chosen as coarse as possible targeting a total compute time of at most 48 hours for 20,000 iterations. We compare this scheme with a finer partitioning along the vertical axis based on 1000 intervals, which results in halving the compute time per iteration.

\begin{figure}
    \centering
    \includegraphics[width=.9\textwidth]{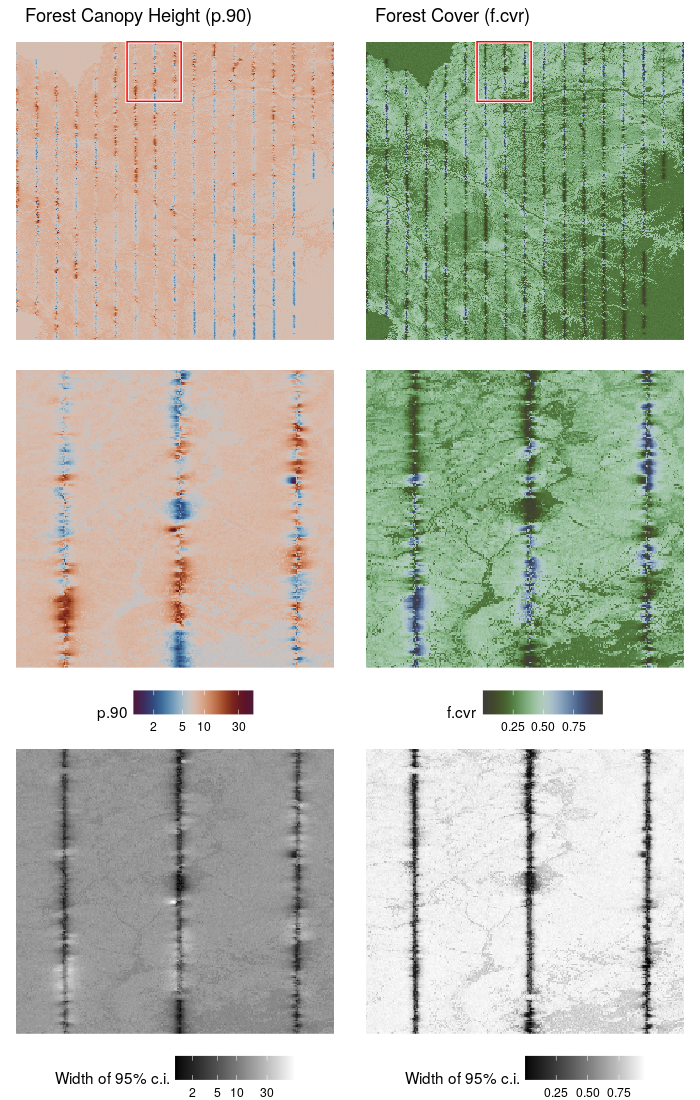}
    \caption{Predictions of \textit{p.90} and \textit{f.cvr} and uncertainty quantification in terms of the width of 95\% credible intervals.}
    \label{fig:alaska:prediction_map}
\end{figure}

Our patterned setup ensures that only a few matrix inverses have to be computed at each iteration. Given that $\text{p.90} > 0$ and $0 < \text{f.cvr} < 1$, we fit (\ref{eq:gripsregression}) on $\tilde{\text{p.90}} = \log\{\text{p.90}\}$ and analogously take $\tilde{\text{f.cvr}} = \log\{ \text{f.cvr} / (1-\text{f.cvr})\})$.
We ran MCMC for a total of 20,000 iterations, of which we discarded the first 5,000 as burn-in. Predictions on a test set made of 10,000 left-out observations are based on a thinned chain consisting of a total of 500 samples---memory constraints inhibit storage of a significantly larger number of samples---and shown in Table \ref{tab:alaska:prediction}. Full prediction maps over the whole study areas along with uncertainty bands are shown in Figure \ref{fig:alaska:prediction_map}. All predictions and their summaries have been performed after inverse-transforming all posterior samples from MCMC.

Prediction results from the two implementations are reported in Table~\ref{tab:alaska:prediction}. All other figures and tables refer to the coarse partition ($33\times 500$) model. We observe that the coarser partitioning strategy is conservative as it results in relatively large regions including up to 140 spatial locations. Finer partitioning leads to faster computations at a minor cost in terms of predictive accuracy. Finally, Table~\ref{tab:alaska:posterior_summaries} provides marginal summaries of the posterior distribution of the unknown covariance and regression parameters of model (\ref{eq:lmc_regression}). For example, $\lambda_{\text{f.cvr}}$ is the diagonal entry of $\bLambda$ which refers to the \textit{f.cvr} outcome, and $\lambda_{\text{p.90}, \text{f.cvr}}$ is the off-diagonal entry of $\bLambda$. Marginal posterior densities and the full Markov chains are available in the Supplement. We find that \textit{loss} has a negative impact on both outcomes even after accounting for spatial variability via the bivariate latent process. The two margins are positively associated as evidenced by the positive value of the off-diagonal term of $\bLambda$ (their correlation at a distance $\bh = 0$ is estimated to be about $\sim 0.686$).
\begin{table}
\centering
\resizebox{.95\textwidth}{!}{
\begin{tabular}{c}
\begin{tabular}{|r|rrrrr|}
  \hline
 $\btheta$ & $\lambda_{\text{p.90}}$ & $\phi_{\text{p.90}}$ &  $\lambda_{\text{f.cvr}}$ & $\phi_{\text{f.cvr}}$ &  $\lambda_{\text{p.90}, \text{f.cvr}}$ \\ 
  \hline
2.5\% quantile & 0.5085 & 232.5055 & 1.1588 & 357.3680  & 1.0777 \\ 
  mean & 0.5213 & 246.4702  & 1.1798 & 376.1024  & 1.1118 \\ 
  97.5\% quantile & 0.5354 & 259.7343  & 1.2012 & 395.0176  & 1.1488 \\  
   \hline
\end{tabular} \\
\\
\begin{tabular}{|r|rrrrrr|}
  \hline
 \multirow{2}{*}{$\bbeta$ and $\btau^2$} & $\beta_{\text{tc}}$ & $\beta_{\text{tc}}$ & $\beta_{\text{loss}}$ & $\beta_{\text{loss}}$ & \multirow{2}{*}{$\tau^2_{\text{p.90}}$} &  \multirow{2}{*}{$\tau^2_{\text{f.cvr}}$} \\ 
   & \small (on \textit{p.90}) & \small (on \textit{f.cvr}) & \small (on \textit{p.90}) & \small (on \textit{f.cvr}) &  &   \\ 
  \hline
  2.5\% quantile & 0.0032 & 0.0162 & -0.0018 & -0.0192 & 0.0157 & 0.1325\\ 
  mean & 0.0032 & 0.0163 & -0.0017 & -0.0187 & 0.0163 & 0.1430\\ 
  97.5\% quantile & 0.0033 & 0.0164 & -0.0015 & -0.0182 & 0.0171 & 0.1533 \\ 
   \hline   
\end{tabular}
\end{tabular}
}
\caption{Estimated posterior summaries for unknown covariance parameters $\btheta$, regression coefficients $\bbeta$, and measurement errors $\btau^2$ for the \modelname\ regression on the Alaskan LiDAR data set.}
\label{tab:alaska:posterior_summaries}
\end{table}

\begin{table}
    \centering
\begin{tabular}{|r|rr|rr|}
  \hline
  & \multicolumn{2}{c|}{$33\times 500$ partition} & \multicolumn{2}{c|}{$33\times 1000$ partition}\\
   & \textit{p.90} & \textit{f.cvr} & \textit{p.90} & \textit{f.cvr}  \\ 
  \hline
  RMSPE & 1.7045 & 0.1181 & 1.7320 & 0.1197 \\ 
  Covg (95\%) & 0.9396 & 0.9384 & 0.9400 & 0.9401 \\ 
  Time/iteration & \multicolumn{2}{c|}{8.18s} & \multicolumn{2}{c|}{3.98s} \\
  Num. iterations & \multicolumn{2}{c|}{20,000} & \multicolumn{2}{c|}{30,000} \\
  Total time & \multicolumn{2}{c|}{45.46h} & \multicolumn{2}{c|}{33.17h}\\
   \hline   
\end{tabular}
\caption{Prediction results over the test set, and timing, of the two implemented models.}
\label{tab:alaska:prediction}
\end{table}

\section{Discussion}
We have introduced gridding and parameter expansion (\modelname) strategies for resolving inefficiencies in MCMC sampling of spatial regression models and demonstrated its validity as an alternative to methods not based on sampling the latent process a posteriori. Our development of the methodology and the subsequent numerical experiments cogently illustrate the effectiveness of \modelname\ in conducting spatial big data analysis. Our focus on MGPs, Mat\'ern covariances and LMCs, and Gaussian outcomes suggests several possible directions for future research. On one hand, we expect our insight to be applicable more generally to Bayesian estimation of spatial regressions with latent processes (not necessarily MGPs). Additional computational considerations arise when non-Gaussian outcomes are paired with latent GPs. In such settings, binary or count outcomes observed in space may be modeled using latent GPs with long-range dependence. Given the good relative performance of \modelname\ in similar contexts in our analyses, we expect \modelname\ to be readily applicable in multivariate non-Gaussian settings \citep[see, e.g.,][]{melange}.

Other extensions include adapting DAG-based processes to spatial-temporal data analysis and to domains of larger dimension. Challenges specific to spatial-temporal domains include identifying ``neighbors'' over space and time as the scale of associations in space significantly differ from that in time. Dynamic NNGPs \citep{nngp_aoas} resolve this problem by estimating neighbors instead of fixing them. QMGPs have also been implemented \citep{meshedgp}, but efficient estimation of the covariance parameters using MCMC remains challenging. Research may, therefore, focus on efficient Bayesian parametrizations of covariance models alternative to either the Matérn or the LMC \citep[see, e.g.,][]{gneiting2010,genton_ccov,zhangbanerjee20,simpsonetal17,fuglstadetal19}. Extensions of \modelname\ to spatial-temporal domains can circumvent some of these problems and, possibly, provide easier and more effective analysis of spatial-temporal big data.


\begin{acks}[Acknowledgments]
M.P. and D.B.D. have received funding from the European Research Council (ERC) under the European Union’s Horizon 2020 research and innovation programme (grant agreement No 856506), and grant R01ES028804 of the United States National Institutes of Health (NIH). S.B. received funding from the National Science Foundation (NSF) through grants NSF/DMS NSF/DMS 2113778, NSF/DMS 1916349 and NSF/IIS 1562303, from the National Institute of Environmental Health Sciences (NIEHS) through grants R01ES030210 and R01ES027027 and from the National Institute of General Medical Science (NIGMS) through R01GM148761. A.O.F. received funding from the U.S. National Science Foundation (NSF) grants DMS-1916395 and EF-1253225, and NASA Carbon Monitoring System program.
\end{acks}

\begin{center}
    {\Large \textsc{Supplement}} 
\end{center}

\section{Models of the response and Collapsed MCMC}
The two main alternatives to posterior sampling of the latent process in a multivariate regression setting with Gaussian measurement errors are (1) modeling the response vector itself as a multivariate GP; (2) marginalizing the scalable GP and running MCMC on the resulting marginal likelihood. In the former case, the multivariate response itself is modeled as a sparse GP: the original cross-covariance function $\Cov_{\btheta}(\bl, \bl')$ is first replaced by $\bK_{\btheta, \btau}(\bl, \bl') = \Cov_{\btheta}(\bl, \bl') + \delta_{\bl = \bl'}\{\bD\}$, then the resulting full model's precision $\bK_{\btheta, \btau}^{-1}$ is replaced by a sparse counterpart $\tilde{\bK}_{\btheta, \btau}^{-1}$. The precision matrix $\tilde{\bK}_{\btheta, \btau}^{-1}$ and its determinant are computed directly from the DAG \citep{nngp_algos}. Because this model does not include latent effects, it will not require them to be sampled. In practice, the underlying sparse DAG leads to fast likelihood evaluations and the underlying unknown parameter is of small dimension. Because these features lead to high efficiency in practice, we compare our proposals to response models in the main article. Despite the efficiency of response models, models with latent effects are sometimes preferrable because they produce inference closer to the full GP model than the response model \citep{katzfuss_vecchia}. 

When seeking to achieve scalability of a spatial regression model with latent effects, it is well known that mixing time increases with the dimension of the state space \citep[see, e.g.,][]{hastings50}. Therefore, without the innovations we have introduced in the main article, a considerably longer Markov chain will be necessary for MCMC to produce samples that accurately characterize the posterior distribution when the dimension of $\bw$ is large as in big data spatial settings. As an alternative to our proposals, one can update the unknown parameters $\{\bbeta, \btau, \btheta\}$ from the marginal or collapsed Bayesian model obtained by integrating out the latent effects $\bw$:
\begin{equation}\label{eq:bayesian_hierarchical_marginal}
    p(\bbeta, \btau, \btheta \given \by) \propto p(\btau, \btheta) \times N(\bbeta;\;\bmu_{\beta},\bV_{\beta})\times N(\by;\; \bX\bbeta, \Cov_{\btheta} + \bD_n)\;. 
\end{equation}
Sampling from (\ref{eq:bayesian_hierarchical_marginal}) now involves $(\Cov_{\btheta} + \bD_n)^{-1}$ and its determinant, both of which can be computed via the Cholesky decomposition of $\bK_{\btheta} = \Cov_{\btheta} + \bD_n$ or its inverse. Once the Cholesky decomposition has been computed, say $\texttt{chol}(\bK_{\btheta}) = \bL\bF\bL^{\top}$, where $\bL=[l_{ii}]$ is unit lower triangular (with $l_{ii}=1$) and $\bF=\mbox{diag}[f_{ii}]$ is diagonal with $f_{ii}> 0$, then 
\begin{equation}\label{eq:gaussian_log_lik}
\log N(\by;\; \bX\bbeta, \Cov_{\btheta} + \bD_n) = \mbox{constant}     -\frac{1}{2}\sum_{i=1}^n \log\left(f_{ii}\right) - \frac{1}{2}\sum_{i=1}^n \frac{u_i^2}{f_{ii}}\;,
\end{equation}
where $u_i$ is the $i$-th element of the vector $\bu$ that solves the lower-triangular system $\bL\bu = \by - \bX\bbeta$. Given the Cholesky decomposition, computing (\ref{eq:gaussian_log_lik}) is cheap (of $O(nq)$). However, the Cholesky decomposition itself involves $O(n^3q^3)$ floating point operations (flops) and $O(n^2q^2)$ in storage complexity and, therefore, is impracticable for very large values of $nq$.
When using a scalable DAG-based process such as an NNGP or an MGP for $\bw$, one uses  the fact that $(\tilde{\Cov}_{\btheta}^{-1} + \bD_n^{-1})^{-1} = \bD_n^{-1} - \bD_n^{-1} (\tilde{\Cov}_{\btheta}^{-1} + \bD_n^{-1})^{-1} \bD_n^{-1}$ and the matrix determinant lemma, which ultimately imply that computational complexity is driven by $(\tilde{\Cov}^{-1} + \bD_n^{-1})^{-1}$ and its determinant. Since $\Cov_{\btheta}^{-1}$ is sparse and $\bD_n^{-1}$ is diagonal, one can use sparse matrix algebra libraries (e.g. sparse Cholesky) which, however, can impede iterative computations in big data analysis. In fact, while $\Cov_{\btheta}^{-1}$ and its determinant are readily available from the DAG, computing $(\tilde{\Cov}^{-1} + \bD_n^{-1})^{-1}$ and its determinants are additional operations that require a re-analysis of the $\tilde{\Cov}^{-1} + \bD_n^{-1}$ matrix. In practical settings with big spatial data, these additional operations may take orders of magnitude more time than the computation of $\Cov_{\btheta}^{-1}$ and its determinant, negating all the advantages of marginalization. Therefore, although by marginalizing $\bw$ out we expect an increase in ESS, the additional time required for marginalization ultimately negates the possible improvements to overall efficiency.

\section{Compute time complexity of QMGP-GriPS}
We make the simplifying assumption that all $q$ outcomes are observed at all $n$ locations, the spatial domain is $\calD = [0,1]^d$, we fix the grid as $\calS = \left( \frac{1}{S+1},\dots, \frac{S}{S+1} \right) \times \cdots \times \left( \frac{1}{S+1},\dots, \frac{S}{S+1} \right)$, so the number of reference locations is $n_S = S^d$. We partition all the domain axes equally into $J$ intervals, resulting in $J^d$ partitions. For ease of exposition, assume that $S$ is a multiple of $J$ so $\calS$ will be subdivided into $M=J^d$ groups, each of size $N = (S/J)^d$. We then build the LMC model on $k$ independent spatial processes with PS-parametrized cross-covariance $\bK$ which we assume stationary (if not, then one can recover the computational complexity directly from the general discussion in \citealt{meshedgp}). 
Each update of the centered parameters requires to evaluate the following density
\begin{align*}
    p^*(\br_\calS ) &= \prod_{i=1}^M N(\br_{i}; \bH_i \br_{[i]}, \bR_i) \times \prod_{\bl \in \calT} N( \by(\bl) \given \bX(\bl)^\top \bbeta + \bcZ_{\bl} \br_{[\bl]}, \bD + \bSigma_{\bl}),
\end{align*}
where $\bH_i^{(j)} = K_j(i, [i]) K_j^{-1}([i], [i])$, $\bR_i^{(j)} = K_j(i, i) - \bH_i^{(j)} K_j([i], i)$. Of the two terms in the product, the second involves $n$ $q$-variate Gaussian densities, therefore its cost is $O(nq^3)$. As for the first term, the LMC assumption implies
\begin{equation}\label{eq:appx:densityterms}
    \prod_{i=1}^M N(\br_{i}; \bH_i \br_{[i]}, \bR_i) = \prod_{j=1}^k \prod_{i=1}^M  N(\br_{i}^{(j)}; \bH_i^{(j)} \br_{[i]}^{(j)}, \bR_i^{(j)}).
\end{equation}
With the QMGP assumption in additions to the others made in this section, $\bH_i^{(j)}$ has $N$ rows and $dN$ or fewer columns; computing it has a cost of $O(d^3 N^3)$. This implies an overall cost of $O(M kd^3 N^3)$. With a stationary $K_j$, we recognize that a QMGP model with gridded $\calS$ shows a redundancy in the $M$ terms, meaning that $\bH_i^{(j)} = \bH_{i^*}^{(j)}$ for some $i^*$. This means that rather than $M$ terms in the product, one needs to compute a much smaller number of unique $\bH_i^{(j)}$ terms. This is referred to as a caching property in \cite{meshedgp}. The minimum number of terms that must be computed is 4 in a gridded QMGP with $d=2$; the maximum depends on the interplay between $\calS$ and the partitioning scheme, but both can be designed to maintain the number of unique terms much smaller than $M$. Suppose that the number of unique terms is $\bar{M} \approx O(1)$, then the overall cost of (\ref{eq:appx:densityterms}) is $O(k d^3 N^3)$ which does not depend on the data size. This implies that sampling $\br$ will be the most expensive step in a regression based on \qmgp-GriPS.

The cost of sampling each block $\br_i$ a posteriori is driven by the required computation of its full conditional covariance, which is of size $kN$. Once we consider that we have $M$ blocks, we obtain an overall cost of $O(M k^3 N^3) = O(n k^3 N^2)$. This cost does not depend on the DAG itself, and specifically it does not depend on the size of the parent set for block $i$ but only on the size of the block itself, as in an independent partition model. 

\section{Full conditional updates} \label{app:fullconds}
We outline the full conditional updates of $\bcA$ and $\btau^2$.
If the data are gridded or we sample $\br$ at non-reference locations, full conditionals for $\bcA$ and $\btau$ are available.
Let $\bY$ be the $n \times q$ matrix whose $j$th column is $\bY_j = \by_j-\bX_j \bbeta_j$, i.e. the $j$th outcome at locations $\calT$, minus the static linear predictor for the same outcome. Similarly denote as $\bF$ the $n \times k$ matrix whose $j$th column is the $j$th margin of the $\br$ process. We can then write $\bY = \bF \bcA^{\top} + \bE$ where $\bE$ is a $n \times q$ matrix such that $\text{vec}\{\bE \} \sim N(\bzero, I_n \otimes \bD)$, with $\text{vec}\{ \cdot\}$ the vectorization operator. For outcome $j$, let $\bF_{(j)}$ subset $\bF$ to columns corresponding to the nonzero elements in the $j$th row of $\bcA$, denoted as $\bcA_{j\setminus 0}$. Then, with $\bD = \text{diag}\{ \tau_1^2, \dots, \tau_q^2 \}$ the full conditional distribution of $\tau_j^2$ is $InvGamma(a_j^*, b^*_j)$ whose parameters are
$a_j^* = a + n_j/2$ and $b^*_j = b + \frac{1}{2}\hat{\bE}_j^\top \hat{\bE}_j$, where $\hat{\bE}_j = \bY_{j, \text{obs}} - \bF_{(j),\text{obs}} \bcA_{j\setminus 0}^{\top} $ and $\bY_{j, \text{obs}}$ is the $n_j \times 1$ vector collecting the observed elements of $\bY_j$. Finally, $\bcA_{j\setminus 0}$ has Gaussian full conditional distribution with precision $\bV_{\bcA_j}^{^{(n)}-1} = \bV_{\bcA_j}^{-1} + \bF_{(j),\text{obs}}^{\top} (I_{n_j} \otimes \bD^{-1}) \bF_{(j), \text{obs}} $ and mean $\bm_{\bcA_j}^{(n)} = \bV_{\bcA_j}^{^{(n)}} \bF_{(j),\text{obs}}^{\top} (I_{n_j} \otimes \bD^{-1}) \bY_{j, \text{obs}}$.

\section{Details on implemented models} \label{app:simdetails}
In the univariate (multivariate) synthetic data applications, the prior distributions on $\tau^2$ ($\tau^2_j$, $j=1,\dots,q$, respectively) were chosen as inverse Gamma with parameters $2.01$ and $1$ for all QMGP and NNGP models. Similarly, $\pi(\sigmasq) = \text{Inv.Gamma}(2.01, 1)$ for QMGPs and NNGPs and similarly for each $\sigmasq_j$ in the multivariate case. In GriPS, $\lambda \sim \pi(\lambda) = N_{\lambda > 0}(0, 1)$ and similarly for $(\bcA)_{ii}$ in the multivariate case.
The SPDE-INLA approach was implemented for $\nu=0.5$ following \cite{inlabook}, Chapter 2, setting $\alpha = \nu + 1 = 1.5$. Setting $\nu=1.5$ was unsupported by \texttt{inla.spde2.pcmatern} in \texttt{R-INLA}. The priors on $\sigmasq, \phi$ were such that $Pr( \sigma > 2.5) = 0.01$ and $Pr(\phi < 0.1) = 0.01$. The mesh for INLA was fixed in the univariate case at a $90\times 90$-sized grid, in the multivariate case the grid size was set to $20\times 20$ in order to limit compute times. Smaller meshes are associated with faster computations in INLAs but negatively impact predictions and estimation accuracy (especially about the nugget $\tau^2$). 
As pointed out during review, the approximation in the SPDE approach works in a different way than our proposed methods; the interpretation of parameters are affected both by boundary effects and discretization. In particular, the nugget variance is inflated by subgrid variation not captured by the piece-wise linear basis. 

All models were run in a Ubuntu 20.04 workstation based on a 16-core, 32-thread AMD Ryzen 9 5950X CPU and 128GB of memory; \texttt{R} version 4.0.3 was linked to Intel MKL 2019.5-075 which includes efficient LAPACK \citep{lapack99} and BLAS \citep{blackford2002blas} libraries for linear algebra. All models were run on 12 threads using OpenMP \citep{dagum1998openmp}.
Effective sample size reported in the tables throughout the article was calculated using the \texttt{effectiveSize} function of R package \texttt{coda} \citep{codapackage}.

\section{Additional details on the LiDAR data application}
\begin{figure}[H]
    \centering
    \includegraphics[width=.9\textwidth]{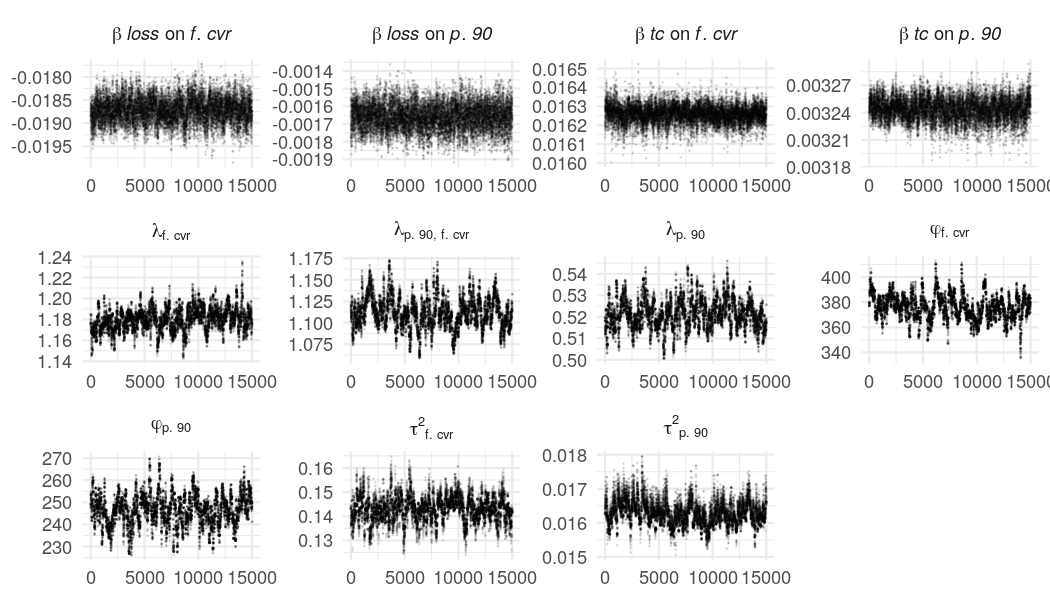}
    \caption{Markov chains for the unknown parameters in the regression model for the LiDAR data.}
    \label{fig:posterior_chains}
\end{figure}

\begin{figure}[H]
    \centering
    \includegraphics[width=.9\textwidth]{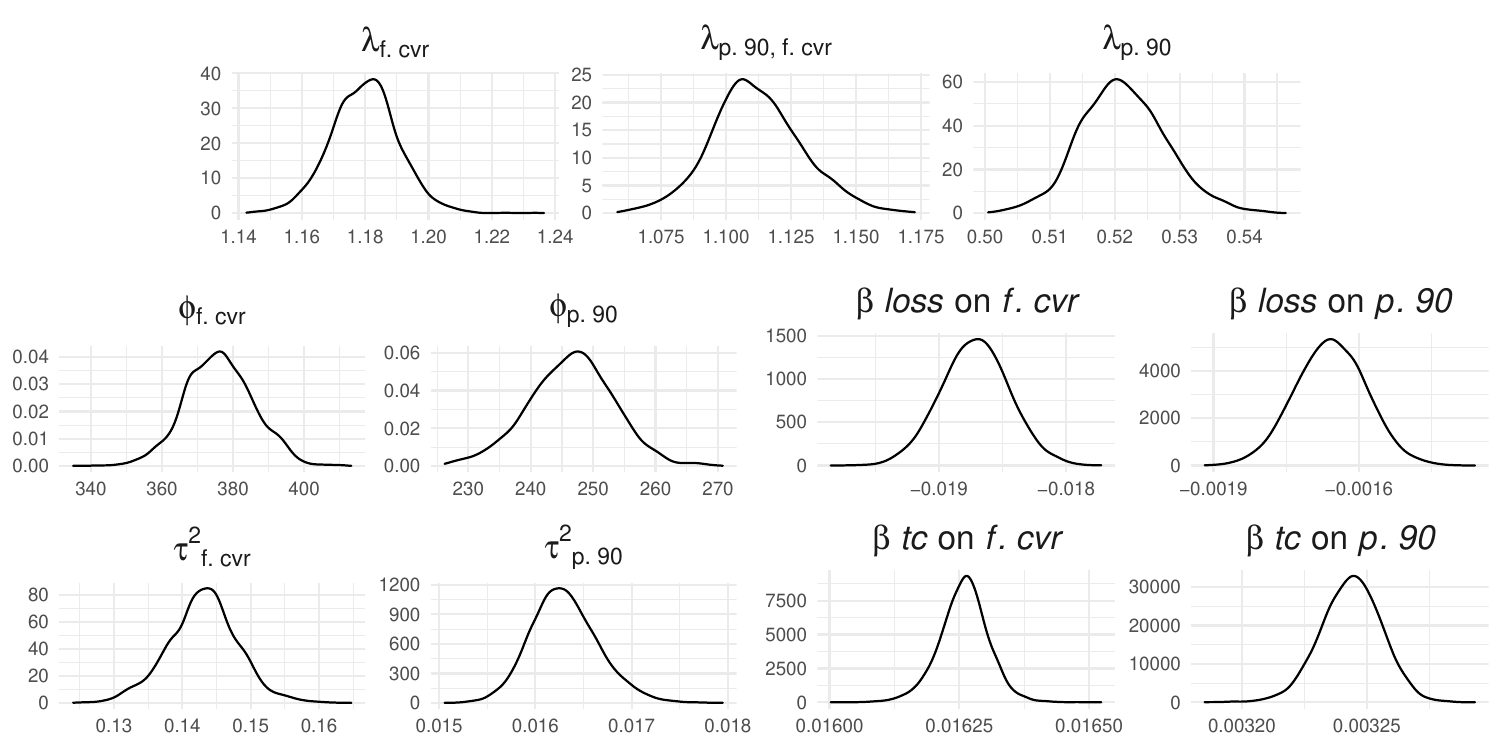}
    \caption{Estimated marginal posterior densities of the covariance parameters $\btheta = \{ \bLambda, \bphi \}$ and regression parameters $\bbeta$ and $\btau^2$.}
    \label{fig:posterior_densities}
\end{figure}

Figures \ref{fig:posterior_chains} and \ref{fig:posterior_densities} add to the posterior summaries reported in the main article.

\bibliographystyle{ba}
\bibliography{biblio.bib}

\end{document}